\documentclass[twocolumn,english,aps,reprint, superscriptaddress,showpacs,longbibliography,showkeys, nofootinbib]{revtex4-1}
\usepackage{amsmath,amssymb,mathtools,bbm,mathrsfs,bm,braket,color,graphicx,comment,xcolor,dsfont,multirow}
\usepackage{hyperref}
\hypersetup{colorlinks,allcolors=blue}
\usepackage[mathscr]{euscript}
\usepackage{subfigure}
\usepackage{multirow}
\usepackage{enumerate}

\begin{document}
	\title{Encoding-Independent Optimization Problem Formulation for Quantum Computing}
	\author{Federico Dominguez} \email{f.dominguez@parityqc.com}

	\affiliation{Parity Quantum Computing Germany GmbH, Ludwigstraße 8, 80539 Munich, Germany}
        
	\author{Josua Unger}
	\affiliation{Parity Quantum Computing GmbH, A-6020 Innsbruck, Austria}
        
	\author{Matthias Traube}
	\affiliation{Parity Quantum Computing Germany GmbH, Ludwigstraße 8, 80539 Munich, Germany}
        
    \author{Barry Mant} 
	\affiliation{Parity Quantum Computing GmbH, A-6020 Innsbruck, Austria}
       
    \author{Christian Ertler}
	\affiliation{Parity Quantum Computing Germany GmbH, Ludwigstraße 8, 80539 Munich, Germany}
        
    \author{Wolfgang Lechner} \email{wolfgang@parityqc.com}
 	\affiliation{Parity Quantum Computing Germany GmbH, Ludwigstraße 8, 80539 Munich, Germany}
  	\affiliation{Parity Quantum Computing GmbH, A-6020 Innsbruck, Austria}
	\affiliation{Institute for Theoretical Physics, University of Innsbruck, A-6020 Innsbruck, Austria}
        
	\date{\today}
	
	\begin{abstract}
		 We present an encoding and hardware-independent formulation of optimization problems for quantum computing. Using  this generalized approach, we present an extensive library of optimization problems and their various derived spin encodings. Common building blocks that serve as a construction kit for building these spin Hamiltonians are identified. This paves the way towards a fully automatic construction of Hamiltonians for arbitrary discrete optimization problems. The presented freedom in the problem formulation is a key step for tailoring optimal spin Hamiltonians for different hardware platforms.

	\end{abstract}
	
	\maketitle
	
	\tableofcontents
	\newpage

	\section{Introduction}
		
		Discrete optimization problems are ubiquitous in almost any field of human endeavor and many of these are known to be NP-hard. 
		The objective in such a discrete optimization problem is to find the minimum of a real valued function $f(v_0,\dots,v_{N-1})$ (the \textit{cost function}) over a set of discrete variables $v_k$. The search space is restricted by hard constraints, which commonly are presented as equalities such as $g(v_0,\dots,v_{N-1}) = 0$ or inequalities as $h(v_0,\dots,v_{N-1})>0$.
		
		Besides using classical heuristics \cite{melnikov2005discrete, dorigo1999ant} and machine learning methods \cite{mazyavkina2021reinforcement} to solve these problems there is also a growing interest in applying quantum computation. One large class for realizing this consists of first encoding the cost function $f$ in a Hamiltonian $H$ such that a subset of eigenvectors (in particular the ground state) of $H$ represents elements in the domain of $f$ and the eigenvalues are the respective values of $f$:
		\begin{equation}
		    H\left|v_0,\dots,v_{N-1}\right\rangle = f(v_0,\dots,v_{N-1}) \left|v_0,\dots,v_{N-1}\right\rangle.
		\end{equation}
		In such an encoding, the ground state of $H$ is the solution of the optimization problem.  
		Having obtained an Hamiltonian reformulation, one can use a variety of quantum algorithms to find the ground state of the Hamiltonian including adiabatic quantum computing \cite{farhi2000quantum} and variational approaches, for instance the quantum/classical hybrid quantum approximate optimization algorithm (QAOA) \cite{farhi2014quantum} or generalizations thereof like the quantum alternating operator ansatz \cite{hadfield2019quantum}. On the hardware side these algorithms can run on gate-based quantum computers, quantum annealers or specialized Ising-machines \cite{mohseni2022ising}.
  
         In the current literature, almost all Hamiltonians for optimization are formulated as Quadratic Unconstrained Binary Optimization (QUBO) problems \cite{Kochenberger2014unconstrained}. The success of QUBO reflects the strong hardware limitations, where higher-than-quadratic interactions are not available. Moreover, quantum algorithms with dynamical implementation of hard constraints \cite{Hen_driver_2016,Hen_quantum_2016} require driver terms that can be difficult to design and also to implement on quantum computers. Hence, hard constraints are usually included as energy penalizations of QUBO Hamiltonians. The prevalence of QUBO has also increased the popularity of one-hot encoding, a particular way of mapping (discrete) variables to eigenvalues of spin operators \cite{lucas2014ising}, since this encoding allows for Hamiltonians with low-order interactions which is especially appropriate for QUBO problems. 
        
        However, compelling alternatives to QUBO and one-hot encoding have been proposed in recent years. A growing number of platforms are exploring high-order interactions \cite{wilkinson2020superconducting,dlaska2022quantum,glaser2022controlled,Chancellor2017circuit, schondorf2019nonpairwise,Menke2021automated,menke2022demonstration,Lu2019global,Pelegri_2022high}, while the Parity architecture \cite{ender2021compiler,lechner2020quantum,fellner2022universal} (a generalization of the LHZ architecture \cite{Lechner2015quantum}) allows the mapping of arbitrary-order interactions to qubits that require only local connectivity. The dynamical implementation of constraints has also been further investigated \cite{hadfield2019quantum,hadfield2017quantum,zhu2022multi,Fuchs2022constraint}, including the design of approximate drivers \cite{sawaya2022encoding} and compilation of constrained problems within the Parity architecture \cite{drieb2021encoding}. Moreover, experimental results have shown that alternative encodings outperform the traditional one-hot approach \cite{chancellor2019domain,jie2021performance,tamura2021performance,sawaya2020resource}. It is clear that alternative formulations for the Hamiltonians need to be explored further, but when the Hamiltonian has been expressed in QUBO using one-hot encoding, it is not trivial to switch to other formulations. Automatic tools to explore different formulations would therefore be highly beneficial. 
        
        We present here a library of problems intended to facilitate the Hamiltonian formulation beyond QUBO and one-hot encoding. Common problems in the literature are revisited and reformulated using encoding-independent formulations, meaning that they can be encoded trivially using any spin encoding. The possible constraints of the problem are also identified and presented separately from the cost function, so the dynamic implementation of the constraints can also be easily explored. Encoding-independent formulations have also been suggested recently \cite{sawaya2022encoding} as an intermediate representation stage of the problems, and exemplified with some use cases. Here, we extend this approach to more than 20 problems and provide a summary of the most popular encodings that can be used. Two additional subgoals are addressed in this library:

		\begin{itemize}
		    \item \textit{Meta parameters/choices}:
		    We present and review the most important choices that can be made in the process of mapping the optimization problem in a mathematical formulation to a spin Hamiltonian. This mainly includes the encodings, which can greatly influence key characteristics important for the computational cost and performance of the optimization, but also free meta parameters or the use of auxiliary variables. All these degrees of freedom are ultimately a consequence of the fact that the optimal solution is typically encoded  only in the ground state and other low-energy eigenstates encode good approximations to the optimal solution. Furthermore, it can be convenient to make approximations so that the solution corresponds not exactly to the ground state but another low-energy state as for example reported in \cite{montanez2022unbalanced}.
		    \item \textit{(partial) automation}:
            Usually, each problem needs to be evaluated individually. The resulting cost functions are not necessarily unique and there is no known trivial way of automatically creating $H$. 
		    By providing a collection of building blocks of cost functions and heuristics for selecting parameters, the creation of the cost function and constraints can be assisted. This enables a general representation of problems in an encoding-independent way and parts of the parameter selection and performance analysis can be conducted in this intermediate stage.

		\end{itemize}

		In practice, many optimization problems are not purely discrete but involve real-valued parameters and variables. Thus, also the encoding of real-valued problems to discrete optimization problems (discretization) as an intermediate step is discussed in Sec.~\ref{sec:real_vars}.

        The document is structured as follows. After introducing the notation used throughout the text in 
        Sec.~\ref{sec:Misc}, we present a list of encodings in Sec.~\ref{sec:enc_lib}. Sec.~\ref{sec:using-lib} then functions as a manual on how to bring the optimization problems contained in this document into a form that can be solved with a quantum computer and two explicit examples are used as illustration. Sec.~\ref{sec:prob_lib} contains a library of optimization problems which are classified into several categories and Sec.~\ref{sec:bb} lists building blocks used in the formulation of these problems. The building blocks are also useful to handle many further problems. Finally, in Sec.~\ref{sec:conclusion} we summarize the results and give an outlook on future projects.

		\section{Definitions and Notation}{\label{sec:Misc}}
	
	In this section we give some basic definitions and settle the notation we will use throughout the whole text.
	
	A \textit{discrete set} of real numbers is a countable subset ${U\subset \mathbb{R}}$ not having an accumulation point. Discrete sets will be denoted by uppercase latin letters, except the letter $G$, which we reserve for graphs. A \textit{discrete variable} is a variable ranging over a discrete set $R$. Discrete variables will be represented by lowercase latin letters, mostly $v$ or $w$. The discrete set $R$ a discrete variable $v$ takes values in, is called its \textit{range}. Elements of $R$ are denoted by lowercase greek letters. 
	
	If a discrete variable has range given by ${\left\lbrace 0,1\right\rbrace}$ we call it \textit{binary} or \textit{boolean}. Binary variables will be denoted by the letter $x$. Similarly, a variable with range ${\left\lbrace -1, 1\right\rbrace}$ will be called a \textit{spin variable} and the letter $s$ will be reserved for spin variables. There is an invertible mapping from a binary variable $x$ to a spin variable $s$:
	\begin{equation}\label{eq: bin to Ising}
	    x\mapsto s\coloneqq 2x-1\, .
	\end{equation}
 For a variable $v$ with range $R$ we define the \textit{value indicator function} to be 
	\begin{equation}\label{eq:value_indicator}
		    \delta_{v}^{\alpha}=\begin{cases}
		    1 \qquad \mathrm{if} \; v=\alpha
		    \\ 0 \qquad \mathrm{if} \; v\neq\alpha,
		    \end{cases}
		\end{equation}
	where $\alpha\in R$.
	
	We will also consider optimization problems for continuous variables. A variable $v$ will be called \textit{continuous}, if its range is given by $\mathbb{R}^d$ for some ${d\in \mathbb{Z}_{>0}}$. 
	
	An \textit{optimization problem $O$} is a triple $(V,f,C)$, where
	\begin{enumerate}
	    \item $V\coloneqq \left\lbrace v_i\right\rbrace_{i=0,\cdots, N-1}$ is a finite set of variables. 
	    \item $f\coloneqq f(v_0, \cdots, v_{N-1})$ is a real valued function, called \textit{objective} or \textit{cost function}.
	    \item $C=\left\lbrace C_i\right\rbrace_{i=0,\cdots, l}$ is a finite set of \textit{constraints} $C_i$. A constraint $C$ is either an equation 
	    \begin{equation}
	        c(v_0,\cdots, v_{N-1})=k
	    \end{equation}
	    for some ${k\in \mathbb{R}}$ and a real valued function $c(v_0,\cdots, v_{N-1})$, or it is an inequality
	    \begin{equation}
	        c(v_0,\cdots, v_{N-1})\leq k\, .
	    \end{equation}
	\end{enumerate}
	
	The goal for an optimization problem ${O=(V,f,C)}$ is to find an extreme value $y_{ex}$ of $f$, such that all of the constraints are satisfied at $y_{ex}$. 

	Discrete optimization problems can often be stated in terms of graphs or hypergraphs. A \textit{graph} is a pair ${(V,E)}$, where $V$ is a finite set of \textit{vertices} or \textit{nodes} and ${E\subset \left\lbrace \left\lbrace v_i,v_j\right\rbrace \, \middle|\, v_i\neq v_j\in V\right\rbrace}$ is the set of edges. An element ${\left\lbrace v_i,v_j\right\rbrace \in E}$ is called an \textit{edge} between vertex $v_i$ and vertex $v_j$. Note that a graph defined like this can neither have loops, i.e., edges beginning and ending at the same vertex, nor can there be multiple edges between the same pair of vertices. Given a graph ${G=(V,E)}$, its \textit{adjacency matrix} is the symmetric binary ${|V|\times |V|}$ matrix $A$ with entries
	\begin{equation}
	    \begin{aligned}
	        A_{ij}=\begin{cases}
	        1,\,&\text{if} \left\lbrace v_i,v_j\right\rbrace \in E,\\ 0,\, &\text{else}\, .
	        \end{cases}
	    \end{aligned}
	\end{equation}
	
	A hypergraph is a generalization of a graph in which we allow edges to be adjacent to more than two vertices. That is, a \textit{hypergraph} is a pair $H=(V,E)$, where $V$ is a finite set of vertices and 

    \begin{equation}
        \begin{aligned}
            E\subseteq \bigcup_{i=1}^{|V|} \big\lbrace \lbrace v_{j_1},\cdots, v_{j_i}\rbrace \, |\, v_{j_k}\in V\, &\mathrm{and}\, v_{j_k}\neq v_{j_{\ell}}\, 
            \\  &  \forall \, k,\ell=1,\cdots i \big\rbrace
        \end{aligned}
    \end{equation}
    is the set of \textit{hyperedges}. 

    Throughout the whole paper we reserve the word \textit{qubit} for actual physical qubits. To get from an encoding-independent Hamiltonian to a quantum program, binary or spin variables become Pauli-z-matrices which act on the corresponding qubits. 
    
    \section{Encodings library}\label{sec:enc_lib}
		\noindent For many important problems, the cost function and the problem constraints can be represented in terms of two fundamental building blocks: the value of the integer variable $v$ and the value indicator $\delta_v^{\alpha}$ defined in Eq.~\eqref{eq:value_indicator}.
		When expressed in terms of these building blocks, Hamiltonians are more compact and recurring terms can be identified across many different problems. Moreover, quantum operators are not present at this stage: an encoding-independent Hamiltonian is just a cost function in terms of discrete variables, which eases access to quantum optimization to a wider audience. The encoding of the variables and the choice of quantum algorithms can be done at later stages. 
		
		The representation of the building blocks in terms of Ising operators depends on the encoding we choose. An encoding is a function that associates eigenvectors of the $\sigma_z$ operator with specific values of a discrete variable $v$:
		\begin{equation}
		    | s_0,\dots,s_{N-1} \rangle \rightarrow v, \quad s_i = \pm 1,
		\end{equation}
		where the spin variables $s_i$ are the eigenvalues of $\sigma_z^{(i)}$ operators. The encodings are also usually defined in terms of binary variables $x_i = 0,1$, which are related to Ising variables $s_i=-1, 1$ according to Eq.~\eqref{eq: bin to Ising}.

\begin{figure*}
	        \centering
	        \includegraphics[width=\textwidth]{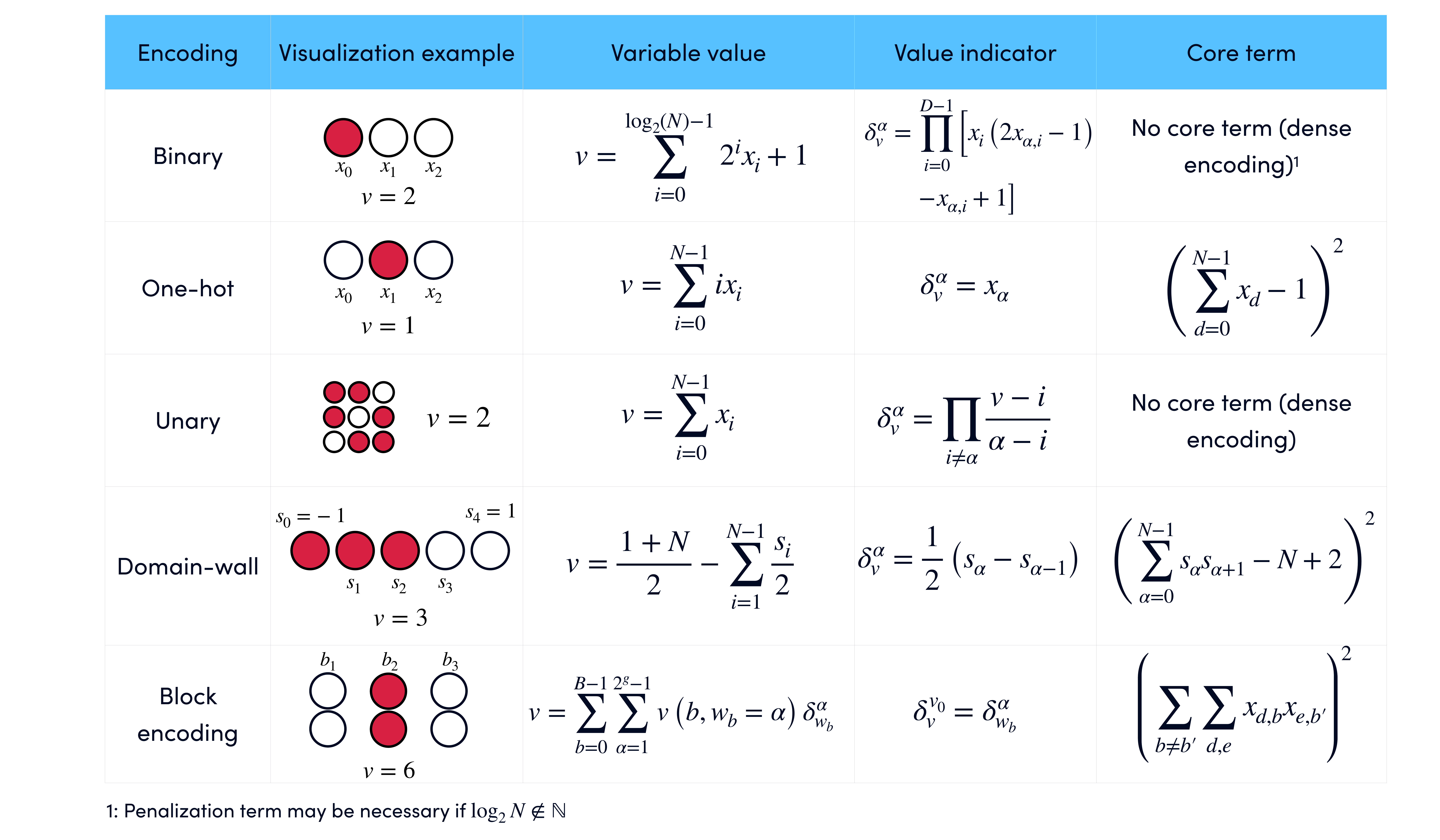}
	        \caption{Summary of popular encodings of discrete variables in terms of binary $x_i$ or Ising $s_i$ variables. Each encoding has a particular representation of the value of the variable $v$ and the value indicator $\delta_v^{\alpha}$. If every quantum state in the register represents a valid value of $v$, the encoding is called \textit{dense}. On the contrary, \textit{sparse encodings} must include a core term in the Hamiltonian in order to represent a value of $v$. Sparse encodings have simpler representations of the value indicators than dense encodings, but the core term implementation demands extra interaction terms between the qubits. The cost of a Hamiltonian in terms of number of qubits and interaction terms strongly depends on the chosen encoding, and should be evaluated for every specific problem.}
	        \label{fig:encodings}
	    \end{figure*}

		A summary of encodings is presented in Fig.~\ref{fig:encodings}. Some encodings are \textit{dense}, in the sense that every quantum state  $| s_0,\dots,s_{N-1} \rangle$ encodes some value of the variable $v$. Other encodings are \textit{sparse}, because only a subset of the possible quantum states are valid states. The valid subset is generated by adding a core term in the Hamiltonian for every sparsely encoded variable. In general, dense encodings require fewer qubits, but sparse encodings have simpler expressions for the value indicator $\delta_{v}^{\alpha}$, and are therefore favorable for avoiding higher-order interactions. This is because $\delta_{v}^{\alpha}$ needs to check a smaller number of qubit states to know whether the variable $v$ has value $\alpha$ or not, whereas dense encodings need to know the state of every qubit in the register~\cite{sawaya2020resource,sawaya2022encoding}. 
		
		\subsection{Binary encoding} 
		\noindent Binary encoding uses the binary representation for encoding integer variables. Given an integer variable ${v \in \left[1,2^D\right]}$, we use $D$ binary variables $x_i\in \lbrace 0,1 \rbrace$ to represent $v$:
		
		\begin{equation}\label{eq:variable_bin}
		    v = \sum_{i=0}^{D-1} 2^i x_i + 1.
		\end{equation}
		
		The value indicator $\delta_v^{\alpha}$ can be written using the generic expression
		\begin{equation}
		    \delta_v^{\alpha} = \prod_{i\neq \alpha} \frac{v-i}{\alpha-i},
		\end{equation}
		which is valid for every encoding. The expression for $\delta_v^{\alpha}$ in terms of boolean variables $x_i$ can be written using that the value of $\alpha$ is codified in the bitstring $(x_{\alpha,0}\dots,x_{\alpha,D-1})$. The value indicator $\delta_{v}^{\alpha}$ checks if the $D$ binary variables $x_i$ are equal to $x_{\alpha,i}$ to know if the variable $v$ has the value $\alpha$ or not. We note that
		\begin{equation}
		    x_{1}\left(2x_2-1 \right)-x_2+1 = \begin{cases}
		    1 \qquad \mathrm{if} \; x_1=x_2
		    \\0 \qquad \mathrm{if} \; x_1\neq x_2
		    \end{cases}
		\end{equation}
		and so we write
		\begin{equation}\label{eq:indicator_bin}
    		\begin{aligned}
    		    \delta_v^{\alpha} &= \prod_{i=0}^{D-1} \left[x_{i}\left(2x_{\alpha,i}-1 \right)-x_{\alpha,i}+1\right]
    		    \\ &= \prod_{i=0}^{D-1} s_{i}\left(x_{\alpha,i} -\frac{1}{2}\right)+\frac{1}{2},  
    		\end{aligned}
		\end{equation}
		where
		\begin{equation}\label{eq:bin2ising}
		    s_{i} = 2x_{i}-1
		\end{equation}		
		are the corresponding Ising variables.  Thus, the maximum order of the interaction terms in $\delta_v^{\alpha}$ scales linearly with $D$ and there are $\binom{D}{k}$ terms of order $k$. The total number of terms is $\sum_k\binom{D}{k}=2^D$. Since $D$ binary variables encode a $2^D$-value variable, then the number of interaction terms needed for a value indicator in binary encoding scales linearly with the size of the variable.

		If $v\in\lbrace 1,\dots,K \rbrace$ with $2^D <K<2^{D+1} ,\; D\in\mathbb{N}$, then we require $D+1$ binary variables to represent $v$ and we will have $2^{D+1}-K$ invalid quantum states that do not represent any value of the variable $v$. The set of invalid states is
		\begin{equation}
		    R = \left\{ |\mathbf{x}\rangle \left| \sum_{i=0}^{D} 2^ix_i +1 > K \right. \right\}. 
		\end{equation}
		For rejecting quantum states in $R$, we force $v\leq K$, which can be accomplished by adding a core term $H_{\mathrm{core}}$ in the Hamiltonian
		\begin{equation}
		    H_{\mathrm{core}}=\sum^{2^{D+1}}_{\alpha=K+1} \delta_v^{\alpha}, 
		\end{equation}
		or imposing the sum constraint
		\begin{equation}
		    c_{\mathrm{core}}=\sum_{\alpha=K+1}^{2^{D+1}} \delta_v^{\alpha} = 0 
		\end{equation}
		The core term penalizes any state that represents an invalid value for variable $v$. Because core terms impose an additional energy scale, the performance can reduce when $K\neq 2^D,\;D\in \mathbb{N}$. In some cases, such as the Knapsack problem, penalties for invalid states can be included in the cost function, so there is no need to add a core term or constraints~\cite{tamura2021performance} (see also Sec.~\ref{bb:ineq}). 

        When encoding variables which can take on negative values as well, e.g. ${v \in \{-K, -K+1, ..., K'-1, K' \}}$, in classical computing one often uses an extra bit that encodes the sign of the value. However, this might not be the best option for our purposes because one spin flip could then change the value substantially and we do not assume full fault tolerance. There is a more suitable encoding of negative values. Let us consider, e.g., the binary encoding. Instead of the usual encoding, we can simply shift the values
        \begin{equation}
            v   = \sum_{i=0}^{D-1} 2^i x_i + 1 - K ,
        \end{equation}
		where $2^{D-1} < K+K' \leq 2^D$. The expression for the value indicator functions stays the same, only the encoding of the value $\alpha$ has to be adjusted. An additional advantage compared to using  the sign bit is that ranges not symmetrical around zero $(K \neq K')$ can be encoded more efficiently. The same approach of shifting the variable by $-K$ can also be used for the other encodings.

		\subsection{Gray encoding}
		\noindent In binary representation, a single spin flip can lead to a sharp change in the value of $v$, for example $|1000\rangle$ codifies $v=9$ while $|0000\rangle$ codifies $v=1$. To avoid this, Gray encoding reorders the binary representation in a way that two consecutive values of $v$ always differ in a single spin flip. 
		If we line up the potential values $v \in [1, 2^D]$ of an integer variable in a vertical sequence, this encoding in $D$ boolean variables can be described as follows: on the $i$-th boolean variable (which is the $i$-th column from the right) the sequence
		starts with $2^{i-1}$ zeros and continues with an alternating sequence of $2^i$ $1$s and $2^i$ $0$s.
        As an example consider
        \begin{align*}
            1 :& \quad 0000 \\
            2 :& \quad 0001 \\
            3 :& \quad 0011 \\
            4 :& \quad 0010 \\
            5 :& \quad 0110 \\
            6 :& \quad 0111 \\
            ...&            
        \end{align*}
        where $D=4$. On the left-hand side of each row of boolean variables we have the value $v$. If we e.g. track the right-most boolean variable we indeed find that it starts with $2^{1-1}=1$ zero for the first value, $2^{1}$ ones for the second and third value, $2^{1}$ zeros for the third and fourth value and so on.
        
        The value indicator function and the core term remain unchanged except that the
		representation of for example $\alpha$ in the analog of Eq.~\eqref{eq:indicator_bin} also has to be in Gray encoding.
		
		An advantage of this encoding with regard to quantum algorithms is that single spin flips do not cause large changes in the cost function and thus smaller coefficients may be chosen (see discussion in Sec.~\ref{sec:bb}).
		
	    \subsection{One-hot encoding}{\label{ss:onehot}} 
	    \noindent One-hot encoding is a sparse encoding that uses $N$ binary variables $x_{\alpha}$ to encode an $N$-valued variable $v$. The encoding is defined by its variable indicator:
	    \begin{equation} \label{eq:one-hot_delta}
	        \delta_{v}^{\alpha} = x_{\alpha},
	    \end{equation}
	    which means that $v = \alpha$ if  $ x_{\alpha}=1$. 
        The value of $v$ is given by

	    \begin{equation}\label{eq:one-hot_value}
	        v = \sum_{\alpha=0}^{N-1} \alpha x_{\alpha}.
	    \end{equation}
     The physically meaningful quantum states are those with a single qubit in state 1 and so dynamics must be restricted to the subspace defined by
	    \begin{equation}\label{eq:one-hot_core}
	        c_{\mathrm{core}}=\sum_{\alpha=0}^{N-1} x_\alpha -1= 0.
	    \end{equation}
	    One option to impose this sum constraint is to encode it as an energy penalization with a core term in the Hamiltonian: 
	    \begin{equation}\label{eq:one-hot-core-term}
	        H_{\mathrm{core}} = \left(1 - \sum_{\alpha=0}^{N-1} x_{\alpha}\right)^2,
	    \end{equation}
	    which has minimum energy if only one $x_{\alpha}$ is different from zero.
     
     An alternative way to impose the core term can be formulated as follows.
     In the building blocks representation of boolean functions (Sec. \ref{bb:bool}) the representation of the $\mathrm{XOR}(x_0, x_1)$ function is given as $\mathrm{XOR}(x_0, x_1) = x_0 + x_1 - 2 x_0 x_1$. In terms of spin variables this is $(1-s_0 s_1)/2$.
     Likewise, concatenating the XOR function yields 
\begin{align}
\mathrm{XOR}(x_0, \mathrm{XOR}(x_1, ... \mathrm{XOR}(x_{N-2}, x_{N-1})...) & =  \nonumber \\
 \frac{1}{2}\left(1-(-1)^{N-1}\prod_{i=0}^{N-1}  s_i\right). &
\end{align}
This can be used with an extra penalty term that lifts the degeneracy to carry out the one-hot check,
i.e. to check whether exactly one of the $x_i$ is equal to 1. For example, consider $N=3$. Then $\mathrm{XOR}(x_0, \mathrm{XOR}(x_1, x_2)) = 1$ for the desired configurations where exactly one of the $x_i$ is one but also if all three are equal to one. In fact, any odd number of ones will result in a one in general. Defining  
\begin{align}
H_{\mathrm{core}} = & -\mathrm{XOR}(x_0, \mathrm{XOR}(x_1, ... \mathrm{XOR}(x_{N-2}, x_{N-1})...) \nonumber \\
& + \alpha \sum_{i=0}^{N-2}x_i + 1-\alpha
\end{align}
with $0< \alpha < 1$
is then a valid one-hot check which requires $N$ terms when expressed as spin variables \footnote{Note that we only need $N-1$ instead of $N$ terms from the summation since any configuration that makes the XOR chain return one and is not a valid one-hot string has at least three ones in it.}. On the other hand, the standard one-hot check of Eq.~(\ref{eq:one-hot_core}) requires $\mathcal{O}(N^2)$ terms which are at most of quadratic order.
	    
	    \subsection{Domain-wall encoding} 
	    \noindent This encoding uses the position of a domain wall in an Ising chain to codify values of a variable $v$. If the endpoints of an $N+1$ spin chain are fixed in opposite states, there must be at least one domain wall in the chain. Since the energy of a ferromagnetic Ising chain depends only on the number of domain walls it has and not on where they are located, an $N+1$ spin chain with fixed opposite endpoints has $N$ possible ground states, depending on the position of the single domain wall.
	    
	    The codification of a variable $v=1,\dots,N$ using domain wall encoding requires the core Hamiltonian~\cite{chancellor2019domain}:
	    \begin{equation}
	         H_{\mathrm{core}} = - \left[ -s_1 + \sum_{\alpha=1}^{N-2} s_{\alpha}s_{\alpha+1} +s_{N-1}\right].
	    \end{equation}
	    Since the fixed endpoints of the chain do not need a spin representation ($s_0 = -1$ and $s_{N} = 1$), $N-1$ Ising variables $\lbrace s_i\rbrace_{i=1}^{N-1} $ are sufficient for encoding a variable of $N$ values. The minimum energy of $H_{\mathrm{core}}$ is $2-N$, so the core term can be alternatively encoded as a sum constraint:
	    \begin{equation}
	        c_{\mathrm{core}} = - \left[ -s_1 + \sum_{\alpha=1}^{N-2} s_{\alpha}s_{\alpha+1} +s_{N-1}\right] = 2-N.
	    \end{equation}
	    
	    The variable indicator corroborates if there is a domain wall in the position $\alpha$:
	    \begin{equation}
	        \delta_{v}^{\alpha} = \frac{1}{2}\left(s_{\alpha} - s_{\alpha-1} \right),
	    \end{equation}
	    where $s_0\equiv -1 $ and  $s_N\equiv 1$, and the variable $v$ can be written as
	    \begin{equation}
    	    \begin{aligned}
    	        v &= \sum_{\alpha=1}^{N} \alpha \delta_{v}^{\alpha}
    	        \\ &= \frac{1}{2}\left[ (1+N) - \sum_{i=1}^{N-1}s_i \right].
    	    \end{aligned}
	    \end{equation}
	    
	    Quantum annealing experiments using domain wall encoding have shown significant improvements in performance compared to one-hot encoding~\cite{jie2021performance}. This is partly because the required search space is smaller ($N-1$ versus $N$ qubits for a variable of $N$ values), but also because domain-wall encoding generates a smoother energy landscape: in one-hot encoding, the minimum Hamming distance between two valid states is two, whereas in domain-wall, this distance is one. This implies that every valid quantum state in one-hot is a local minimum, surrounded by energy barriers generated by the core energy of Eq.~\eqref{eq:one-hot-core-term}. As a consequence, the dynamics in domain-wall encoded problems freeze later in the annealing process, because only one spin-flip is required to pass from one state to the other.
	    
	    \subsection{Unary encoding} 
	    \noindent For this case, the value of $v$ is encoded in the number of binary variables which are equal to one:
	    \begin{equation}
	        v = \sum_{i=0}^{N-1} x_i,
	    \end{equation}
	    so $N-1$ binary variables $x_i$ are needed for encoding an $N$-value variable. Unary encoding does not require a core term because every quantum state is a valid state. However, this encoding is not unique in the sense that each value of $v$ has multiple representations. 
	    
	    A drawback of unary encoding (and every dense encoding) is that it requires information from all binary variables to determine the value of $v$. The value indicator $\delta_{v}^{\alpha}$ is
	    
    	\begin{equation}
		    \delta_{v}^{\alpha} = \prod_{i\neq \alpha} \frac{v-i}{\alpha-i},   
		\end{equation}
	    which involves $2^N$ interaction terms. This scaling is completely unfavorable, so unary encoding may be only convenient for variables that do not require value indicators $\delta_v^{\alpha}$, but only the variable value $v$. 
	    
	    A performance comparison for the Knapsack problem using digital annealers showed that unary encoding can outperform binary and one-hot encoding and requires smaller energy scales~\cite{tamura2021performance}. The reasons for the high performance of unary encoding are still under investigation, but redundancy is believed to play an important role, because it facilitates the annealer to find the ground state. As for domain-wall encoding \cite{berwald2021understanding}, the minimum Hamming distance between two valid states (i.e., the number of spin flips needed to pass from one valid state to another) could also explain the better performance of the unary encoding. 
	    
	    \subsection{Block encodings} \noindent It is also possible to combine different approaches to obtain a balance between sparse and dense encodings~\cite{sawaya2020resource}. Block encodings are based on $B$ blocks, each consisting of $g$ binary variables. Similar to one-hot encoding, the valid states for block encodings are those states where only a single block contains non-zero binary variables. The binary variables in block $b$, $\lbrace x_{b,i} \rbrace_{i=0}^{g-1}$ define a block value $w_b$, using a dense encoding like binary, gray or unary. For example, if $w_{b}$ is encoded using binary, we have
	    \begin{equation}
	        w_{b} = \sum_{i=0}^{g-1} 2^i x_{b,i} + 1.
	    \end{equation}
	    The discrete variable $v$ is defined by the active block $b$ and its corresponding block value $w_{b}$,
	    \begin{equation}
	        v= \sum_{b=0}^{B-1}\sum_{\alpha=1}^{2^g-1} v\left(b, w_b=\alpha\right) \delta_{w_{b}}^{\alpha},
	    \end{equation}
	    where $v\left(b, w_b=\alpha\right)$ is the discrete value associated to the quantum state with active block $b$ and block value $\alpha$, and $\delta_{w_{b}}^{\alpha}$ is a value indicator that only needs to check the value of binary variables in block $b$. For each block, there are $2^{g-1}$ possible values (assuming gray or binary encoding for the block), because the all-zero state is not allowed (otherwise block $b$ is not active). If the block value $w_b$ is encoded using unary, then $g$ values are possible. The expression of $\delta_{w_{b}}^{\alpha}$ depends on the encoding and is the same as the one already presented in the respective encoding section. 
	    
	    The value indicator for the variable $v$ is the corresponding block value indicator. Suppose the discrete value $v_0$ is encoded in the block $b$ with a block variable $w_b=\alpha$:
	    \begin{equation}
	        v_0=v\left(b, w_b=\alpha\right).
	    \end{equation}
	    Then the value indicator $\delta_{v}^{v_0}$ is
	    \begin{equation}
	        \delta_{v}^{v_0} = \delta_{w_b}^{\alpha}.
	    \end{equation}
	    
	    A core term is necessary so that only qubits in a single block can be in the excited state. Defining $t_b=\sum_i x_{i,b}$, the core terms results in
	    \begin{equation}
	        H_{\mathrm{core}}= \sum_{b\neq b'} t_bt_{b'},
	    \end{equation}
	    or, as a sum constraint,
	     \begin{equation}
	        c_{\mathrm{core}} = \sum_{b\neq b'} t_bt_{b'}=0.
	    \end{equation}
	    The minimum value of $H_{\mathrm{core}}$ is zero. If two blocks $b$ and $b'$ have binary variables with values one, then $t_bt_{b'}\neq 0$ and the corresponding eigenstate of $H_{\mathrm{core}}$ is no longer the ground state.  
	    
	    		\section{Parity Architecture}\label{sec:parity}
		\noindent 
  The strong hardware limitations of noisy intermediate-scale Quantum (NISQ) \cite{preskill2018quantum} devices have made sparse encodings (and especially one-hot encoding) the standard approach to problem formulation. This is mainly because the basic building blocks (value and value indicator) are of linear or quadratic order in the spin variables in these encodings.
  The low connectivity of qubit platforms requires Hamiltonians in the QUBO formulation, and high-order interactions are expensive when translated to QUBO~\cite{Kochenberger2014unconstrained}. However, that different choices of encodings can significantly improve the performance of quantum algorithms~\cite{chancellor2019domain,jie2021performance,tamura2021performance,sawaya2020resource}, since a smart choice of encodings can reduce the search space or generate a smoother energy landscape.
  
		One way this difference between encodings manifests itself is in the number of spin flips of  physical qubits needed to change a variable into another valid value~\cite{berwald2021understanding}. If this number is larger than one, there are local minima separated by invalid states penalized with a high cost which can impede the performance of the optimization. On the other hand, such an energy-landscape might offer some protection against errors (see also Ref.~\cite{fellner2022universal} and Ref.~\cite{pastawski2016error}).
		Furthermore, other fundamental aspects of the algorithms, such as circuit depth and energy scales can be greatly improved outside QUBO~\cite{Ender2022modular,drieb2021encoding,Fellner2021benchmarks}, prompting us to look for alternative formulations to the current QUBO-based Hamiltonians. The Parity Architecture is a paradigm for solving quantum optimization problems~\cite{Lechner2015quantum,ender2021compiler} that does not rely on the QUBO formulation, hence allowing a wide number of options for formulating Hamiltonians. The architecture is based on the Parity transformation, which remaps Hamiltonians into a two-dimensional grid that requires only local connectivity of the qubits. The absence of long-range interactions enables high parallelizability of quantum algorithms and eliminates the need for costly and time-consuming SWAP gates, which helps to overcome two of the main obstacles of quantum computing: the limited coherence time of qubits and the poor connectivity of qubits within a quantum register. 
		
		The Parity transformation creates a single Parity qubit for each interaction term in the (original) logical Hamiltonian:
		\begin{equation}
		    J_{i,j,\dots} \sigma_{z}^{(i)}\sigma_{z}^{(j)}\dots \rightarrow J_{i,j,\dots} \sigma_z^{(i,j,\dots)},
		\end{equation}
		where the interaction strength $J_{i,j,\dots}$ is now the local field of the Parity qubit $\sigma_z^{(i,j,\dots)}$. This facilitates addressing high-order interactions and frees the problem formulation from the QUBO approach. The equivalence between the original logical problem and the Parity-transformed problem is ensured by adding three-body and four-body constraints and placing them on a two dimensional grid such that only neighboring qubits are involved in the constraints. The mapping of a logical problem into the regular grid of a Parity chip can for example be realized by the Parity compiler~\cite{ender2021compiler}. Although the Parity compilation of the problem may require a larger number of physical qubits, the locality of interactions on the grid allows for higher parallelizability of quantum algorithms. This allows constant depth algorithms~\cite{lechner2020quantum,unger2022low} to be implemented with a smaller number of gates~\cite{Fellner2021benchmarks}.  
		
\begin{figure}
            \centering
            \includegraphics[width=\linewidth]{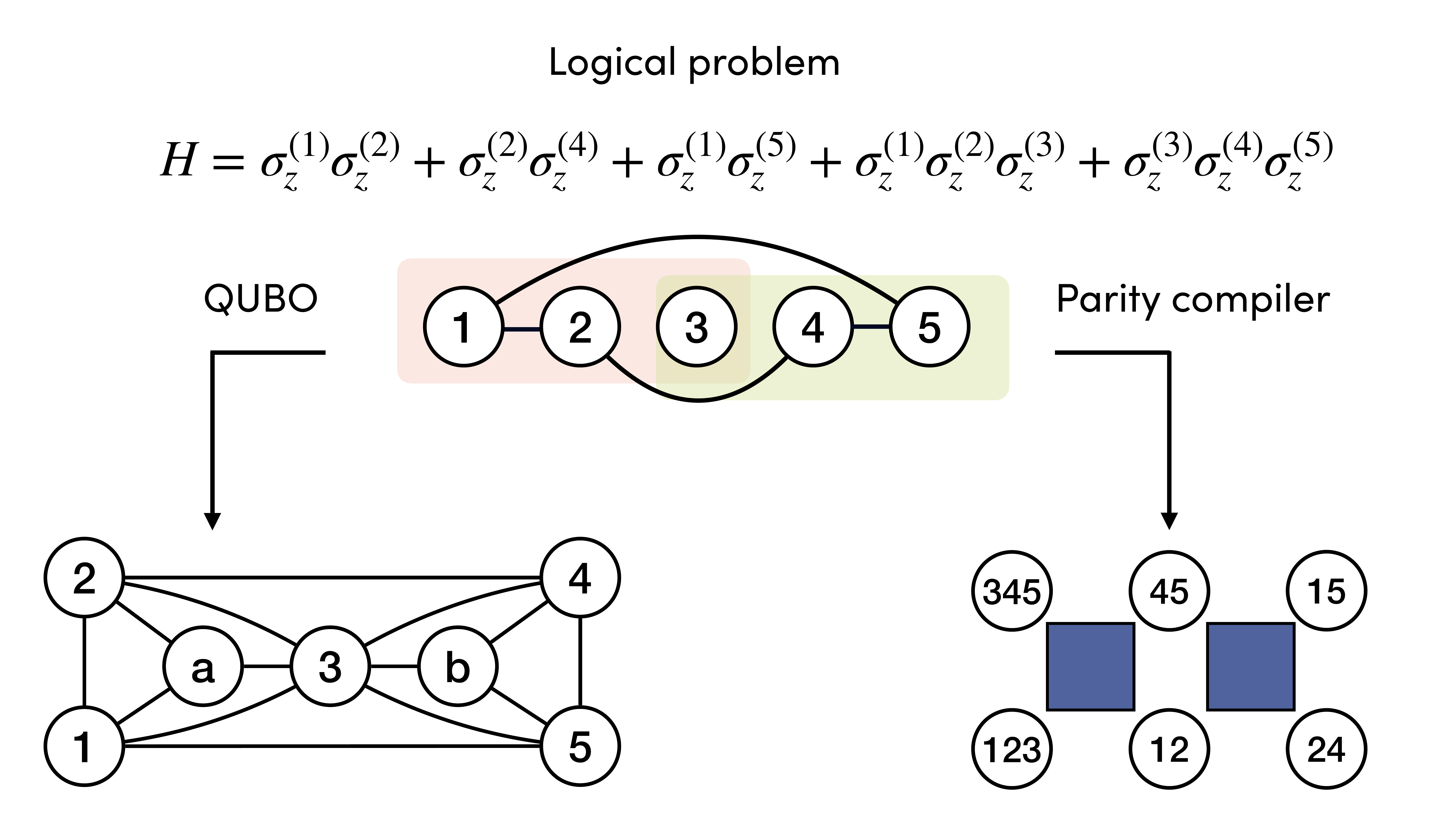}
            \caption{Example toy problem $H$ involving high-order interactions that shows how the Parity architecture straightforwardly handles Hamiltonians beyond the QUBO formulation. The problem is represented by the hypergraph at the top of the figure. When decomposed into QUBO form, it requires 7 qubits and 14 two-body interactions, plus additional qubit overhead depending on the embedding. In contrast, the Parity compiler can remap this problem into a two-dimensional grid that requires six qubits with local fields and four-body interactions, represented by the blue squares. No additional embedding is necessary if the hardware is designed for the Parity Architecture}
            \label{fig:parity_qubo}
        \end{figure}

		The following toy example, summarized in Fig.~\ref{fig:parity_qubo}, shows how a Parity-transformed Hamiltonian can be solved using a smaller number of qubits when the original Hamiltonian has high-order interactions. Given the logical Hamiltonian
		\begin{equation}
		    \begin{aligned}
		         H = &\sigma_{z}^{(1)}\sigma_{z}^{(2)}+\sigma_{z}^{(2)}\sigma_{z}^{(4)}
		         \\&+\sigma_{z}^{(1)}\sigma_{z}^{(5)}+\sigma_{z}^{(1)}\sigma_{z}^{(2)}\sigma_{z}^{(3)}+\sigma_{z}^{(3)}\sigma_{z}^{(4)}\sigma_{z}^{(5)},
		    \end{aligned}
		\end{equation}
        the corresponding QUBO formulation requires $5+2=7$ qubits, including two ancilla qubits for decomposing the three-body interactions, and the total number of two-body interactions is 14. The embedding of the QUBO problem in the quantum hardware may require additional qubits and interactions depending on the chosen architecture. Instead, the Parity-transformed Hamiltonian only consists of six Parity qubits with local fields and two four-body interactions between close neighbors.
        
        It is not yet clear what the best Hamiltonian representation is for an optimization problem. The answer will probably depend strongly on the particular use case we want to solve, and will take into account not only the number of qubits needed, but also the smoothness of the energy landscape, which has a direct impact on the performance of quantum algorithms \cite{king2019quantum}. The Parity architecture allows us to explore formulations beyond the standard QUBO and this library aims to facilitate the exploration of new Hamiltonian formulations. 
        
   \section{Encoding constraints} \label{ss: constraints}
	        
	        \noindent Up to this point, we have presented the possible encodings of the discrete variables in terms of binary variables. In this section, we assume that the encodings of the variables have already been chosen and explain how to implement the hard constraints associated with the problem. Hard constraints $c(v_1,\dots,v_N)=K$ often appear in optimization problems, limiting the search space and making problems even more difficult to solve. We consider polynomial constraints of the form
	        \begin{equation}\label{eq:pol_constraint}
	            \begin{aligned}
	                c(v_1,\dots,v_N) = & \sum g_i v_i
	                \\& + \sum g_{i,j}v_iv_j + \sum g_{i,j,k} v_iv_jv_k + \dots,
	            \end{aligned}
	        \end{equation}
	        which remain polynomial after replacing the discrete variables $v_i$ with any encoding. The coefficients $g_i,g_{i,j},\dots$ depend on the problem and its constraints. 
         
         In general, constraints can be implemented dynamically~\cite{Hen_driver_2016,Hen_quantum_2016} (exploring only quantum states that satisfy the constraints) or as extra terms $H_{core}$ in the Hamiltonian, such that eigenvectors of $H$ are also eigenvectors of $H_{core}$ and the ground states of $H_{core}$ correspond to elements in the domain of $f$ that satisfy the constraint. Even if the original problem is unconstrained, the use of sparse encodings such as one-hot or domain wall imposes hard constraints on the quantum variables. 
         
         Constraints arising from sparse encodings are usually incorporated as a penalty term $H_{core}$ into the Hamiltonian that penalizes any state outside the desired subspace:
	        \begin{equation}\label{eq:constraint_sq}
	            H_{\mathrm{core}} = A\left[c(v_1,\dots,v_N)-K\right]^2,
	        \end{equation}
	        or
	        \begin{equation}
	            H_{\mathrm{core}} = A\left[c(v_1,\dots,v_N)-K\right],
	        \end{equation}
	        in the special case that $c(x_1,\dots,x_N)\geq K$ is satisfied. The constant $A$ must be large enough to ensure that the ground state of the total Hamiltonian satisfies the constraint, but the implementation of large energy scales lowers the efficiency of quantum algorithms~\cite{Lanthaler_2021} and additionally imposes a technical challenge. Moreover, extra terms in the Hamiltonian imply additional overhead of computational resources, especially for squared terms like in Eq.~\eqref{eq:constraint_sq}. 
	        
	        Quantum algorithms for finding the ground state of Hamiltonians, like QAOA or quantum annealing, require driver terms $U_{\mathrm{drive}}=\exp(-itH_{\mathrm{drive}})$ that spread the initial quantum state to the entire Hilbert space. Dynamical implementation of constraints employs a driver term that only explores the subspace of the Hilbert space that satisfies the constraints. Given an encoded constraint in terms of Ising operators $\sigma_{z}^{(i)}$:,
	        \begin{equation}\label{eq:pol_constraint_ising}
	            \begin{aligned}
	                c(\sigma_{z}^{(1)},\dots,\sigma_{z}^{(N)}) =& \sum g_i \sigma_{z}^{(i)} 
	                \\& + \sum g_{i,j} \sigma_{z}^{(i)} \sigma_{z}^{(j)} + \dots =K,
	            \end{aligned}
	        \end{equation}
	        a driver Hamiltonian $H_{\mathrm{drive}}$ that commutes with $c(\sigma_{z}^{(1)},\dots,\sigma_{z}^{(N)})$ will restrict to the valid subspace provided the initial quantum state satisfies the constraint:
	        \begin{equation}
	            \begin{aligned}
	                c|\psi_0\rangle &= K|\psi_0\rangle
	                \\ U_{\mathrm{drive}}\; c|\psi_0\rangle &= U_{\mathrm{drive}}\;K|\psi_0\rangle
	                \\ c\left[U_{\mathrm{drive}} |\psi_0\rangle\right] &= K\left[U_{\mathrm{drive}} |\psi_0\rangle\right].
	            \end{aligned}
	        \end{equation}
	        
	        \begin{figure}
	            \centering
	            \includegraphics[width=\linewidth]{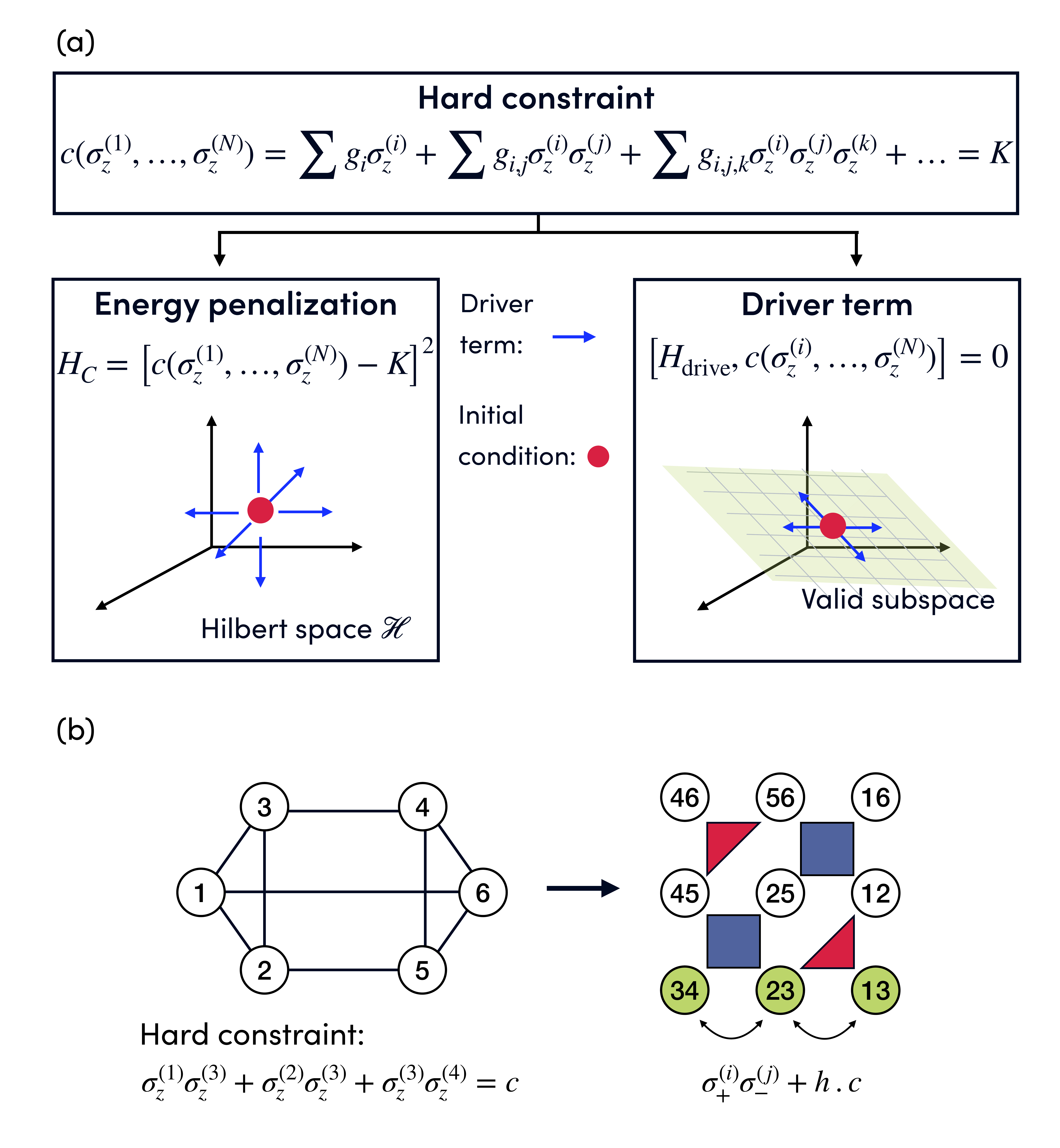}
	            \caption{(a) Decision tree to encode the constraints, which can be implemented as energy penalizations in the Hamiltonian of the problem or dynamically by selecting a driver Hamiltonian that preserves the desired condition. When using energy penalties (left figure), the driver term needs to explore the entire Hilbert space. In contrast, the dynamic implementation of the constraints (right figure) reduces the search space to the subspace satisfying the hard constraint, thus improving the performance of the algorithms. (b) Constrained logical problem (left) and its Parity representation. In the Parity architecture, each term of the polynomial constraint is represented by a single Parity qubit (green qubits), so the polynomial constraints define a subspace in which the total magnetization of the involved qubits is preserved and which can be explored with a exchange driver $\sigma_+\sigma_-+h.c.$ that preserves the total magnetization. Figure originally published in Ref.~\cite{drieb2021encoding}.}
	            \label{fig:enc_const}
	        \end{figure}
	    
	    In general, the construction of constraint-preserving drivers depends on the problem~\cite{Hen_driver_2016,Hen_quantum_2016,sawaya2022encoding,chancellor2019domain, hadfield2017quantum}. Approximate driver terms have been proposed, which admit some degree of leakage and may be easier to construct~\cite{sawaya2022encoding}. Within the Parity Architecture, each term of a polynomial constraint is a single Parity qubit~\cite{Lechner2015quantum,ender2021compiler}. This implies that for the Parity architecture the polynomial constraints are simply the conservation of magnetization between the qubits involved:  
	    \begin{equation}
	        \begin{aligned}
	            \sum_i g_i \sigma_{z}^{(i)} &+ \sum_{i,j}g_{i,j}\sigma_{z}^{(i)}\sigma_{z}^{(j)} + \dots = K
	            \\ & \rightarrow \sum_{\mathbf{u} \in c}g_{\mathbf{u}} \sigma_{z}^{(\mathbf{u})} = K,
	        \end{aligned}
	    \end{equation}
	    where the Parity qubit $\sigma_{z}^{(\mathbf{u})}$ represents the logical qubits product $\sigma_{z}^{(u_1)}\sigma_{z}^{(u_2)}\dots \sigma_{z}^{(u_n)}$ and we sum over all these products that appear in the constraint $c$. A driver Hamiltonian based on exchange (or flip-flop) terms $\sigma_+^{(\mathbf{u})}\sigma_-^{(\mathbf{w})}+h.c.$, summed over all pairs of products $\mathbf u, \mathbf w \in c$, preserves the total magnetization and explores the complete subspace where the constraint is satisfied \cite{drieb2021encoding}. The decision tree for encoding constraints is presented in Fig.~\ref{fig:enc_const}.  
	\section{Encoding-Independent Formulation}\label{sec:using-lib}
	    \noindent The problems given in this library consist of a cost function $f$ and a set of constraints that define a subspace where the function $f$ must be minimized. Both the cost function and the constraints are expressed in terms of discrete variables and we call this \textit{encoding independent formulation}. To generate a Hamiltonian suitable for quantum optimization algorithms, the discrete variables must be encoded in Ising operators, using the encodings presented in Sec.~\ref{sec:enc_lib}. If sparse encodings are chosen for some variables, then the core constraints of these variables must be included in the constraint set of the problem. 

        After encoding the problem, it must be decided whether the constraints (either problem constraints or core constraints) are implemented as energy penalties or dynamically in the driver term (see Sec.~\ref{ss: constraints}). The Hamiltonian of the problem will consist of the encoded cost function $f$ plus the constraints $c_i$ that are implemented as energy penalties:
        \begin{equation}
            H = Af + Bc_1 + C c_2 + \dots,
        \end{equation}
	    where the energy scales $A<B<C,\dots$ are selected such that the ground state of $H$ is also the ground state of each of the terms representing constraints. This guarantees that it is never energetically favorable to violate a constraint, in order to further reduce the cost function. If the energy scales are too different, the performance of the algorithms may deteriorate (in addition to the experimental challenge that this entails) so parameters $A,B,C,\dots$ cannot be set arbitrarily large. 
	    The determination of the optimal energy scales is an important open problem. For some of the problems in the library we provide an estimation of the energy scales (cf. also Sec.~\ref{bb:prio}). 
	    
	    The problems included in the library are classified into four different categories: subsets, partitions, permutations, continuous variables. These categories are defined by the role of the discrete variable and are intended to organize the library and make it easier to find problems, but also to serve as a basis for the formulation of similar use cases. An additional category in Sec.~\ref{ss:other_problems} contains problems that do not fit into the previous categories but may also be important use cases for quantum algorithms. In Sec.~\ref{sec:bb} we include a summary of recurrent encoding-independent building blocks that are used throughout the library and could be useful in formulating new problems.
	    
	    We now present two examples, going from the encoding independent problem to the Ising Hamiltonian and its constraints. This output can already be compiled using the Parity compiler~\cite{ender2021compiler} without the need for reducing the problem to QUBO or additional embedding.  
	    
	Let us illustrate the encoding for two examples, the clustering problem and the traveling salesman problem. 
 
\begin{figure*} 
              \includegraphics[width=\textwidth]{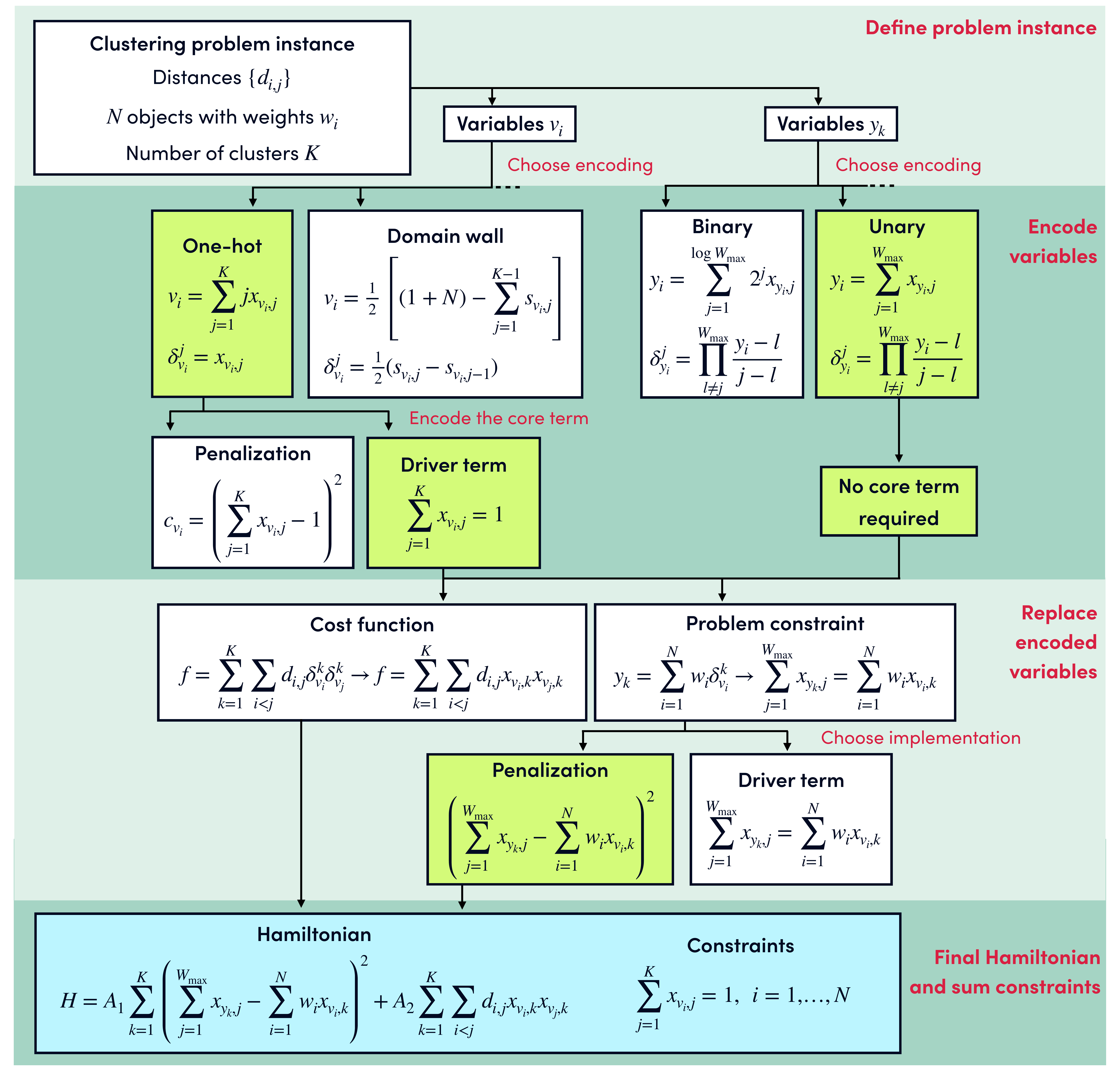}
              \caption{Example decision tree for the clustering problem. The entire process is presented in four different blocks (Define problem, Encode variables, Replace variables and Final Hamiltonian) and the decisions taken are highlighted in green. The formulation of a problem instance requires the discrete variables $v_i$ and $y_k$ to be encoded in terms of qubit operators. In this example, the $v_i$ variables are encoded using the one-hot encoding while for the $y_k$ variables we use the binary encoding. The Hamiltonian is obtained by substituting in the encoding-independent expressions of Eqs.~\eqref{eq:clustering_constraint} and~\eqref{eq:clustering_cost} the discrete variables $v_i,y_k$ and the value indicators $\delta_{v}^{\alpha}$ according to the chosen encoding. Core terms associated to sparse encodings (one-hot in this case) and the problem constraints can be added to the Hamiltonian as energy penalizations or implemented dynamically in the driver term of the quantum algorithm as sum constraints. The final Hamiltonian and the sum constraints can be directly encoded on a Parity chip using the Parity Compiler, without the need to reduce the problem to its QUBO formulation.} \label{fig:clustering_tree} 
            \end{figure*}

 \paragraph{Clustering problem:} 
	        \noindent The clustering problem is a partitioning problem described in Sec.~\ref{subsub: clustering_prob}. The process to go from the encoding-independent Hamiltonian to the spin Hamiltonian that has to be implemented in the quantum computer is outlined in Fig.~\ref{fig:clustering_tree}. A problem instance is defined from the required inputs: the number of clusters $K$ we want to create, $N$ objects with weights $w_i$ and distances $d_{i,j}$ between the objects. Two different types of discrete variables are required, variables $v_i=1,\dots,K$ ($i=1,\dots,N$) indicate to which of the $K$ possible clusters the object $i$ is assigned, and variables $y_j = 1,\dots,W_{\mathrm{max}}$ track the weight in cluster $j$. 
	        
	        The variables $v_i$ and $y_j$ serve different purposes in the formulation of the clustering problem and thus it might be beneficial to use different encodings. For variables $v_i$ only the value indicator $\delta_{v_i}^{\alpha}$ is required, therefore sparse encodings with simple $\delta_{v_i}^{\alpha}$ are recommended. In contrast, only the value of variables $y_j$ is required for the problem formulation, and so dense encodings like binary or Gray will reduce the number of necessary boolean variables. In order to find the spin representation of the problem, it is necessary to choose an encoding and decide if the core term (if there is one) is included as an energy penalization or via a constraint-preserving driver (see Sec.~\ref{ss: constraints}). 
	        
	        Once the encodings have been chosen, the discrete variables are replaced in the cost function and in the constraint associated to the problem, Eqs.~\eqref{eq:clustering_cost} and~\eqref{eq:clustering_constraint}, respectively. The problem constraint can be encoded as an energy penalization in the Hamiltonian or implemented dynamically in the driver term of the quantum algorithm as a sum constraint. The resulting Hamiltonian and constraints are now in terms of spin operators, and can compiled using the Parity compiler to obtain a two-dimensional chip design including only local interactions.  
	        
        \paragraph{Traveling salesperson problem:}
            \noindent The second example we present is the Traveling salesperson (TSP) problem, described in Sec.~\ref{sec:TSP_problem}. The input to this problem is a (usually complete) graph $G$ with a cost $w_{i,j}$ associated with each edge in the graph. We seek a path that visits all nodes of $G$ without visiting the same node twice, and also minimizes the total cost $\sum w_{i,j}$ of the edges of the path. 
            
            This problem requires $N$ variables $v_i=1,\dots,N$, where $N$ is the number of vertices in $G$. Each variable $v_i$ represents a node in the graph, and indicates the stage of the travel at which node $i$ is visited. If $v_i=\alpha$ and $v_j=\alpha+1$, the salesperson travels from node $i$ to node $j$ and so the cost function adds $w_{i,j}$ to the total cost of the travel, as indicated in Eq.~\eqref{eq:tsp_cost}. This formulation requires the constraints of Eq.~\eqref{eq:permutation_constraint} $v_i\neq v_j,\,$ for all$ \,i\neq j$ so the salesperson visits every node once.
            
\begin{figure*} 
              \includegraphics[width=\textwidth]{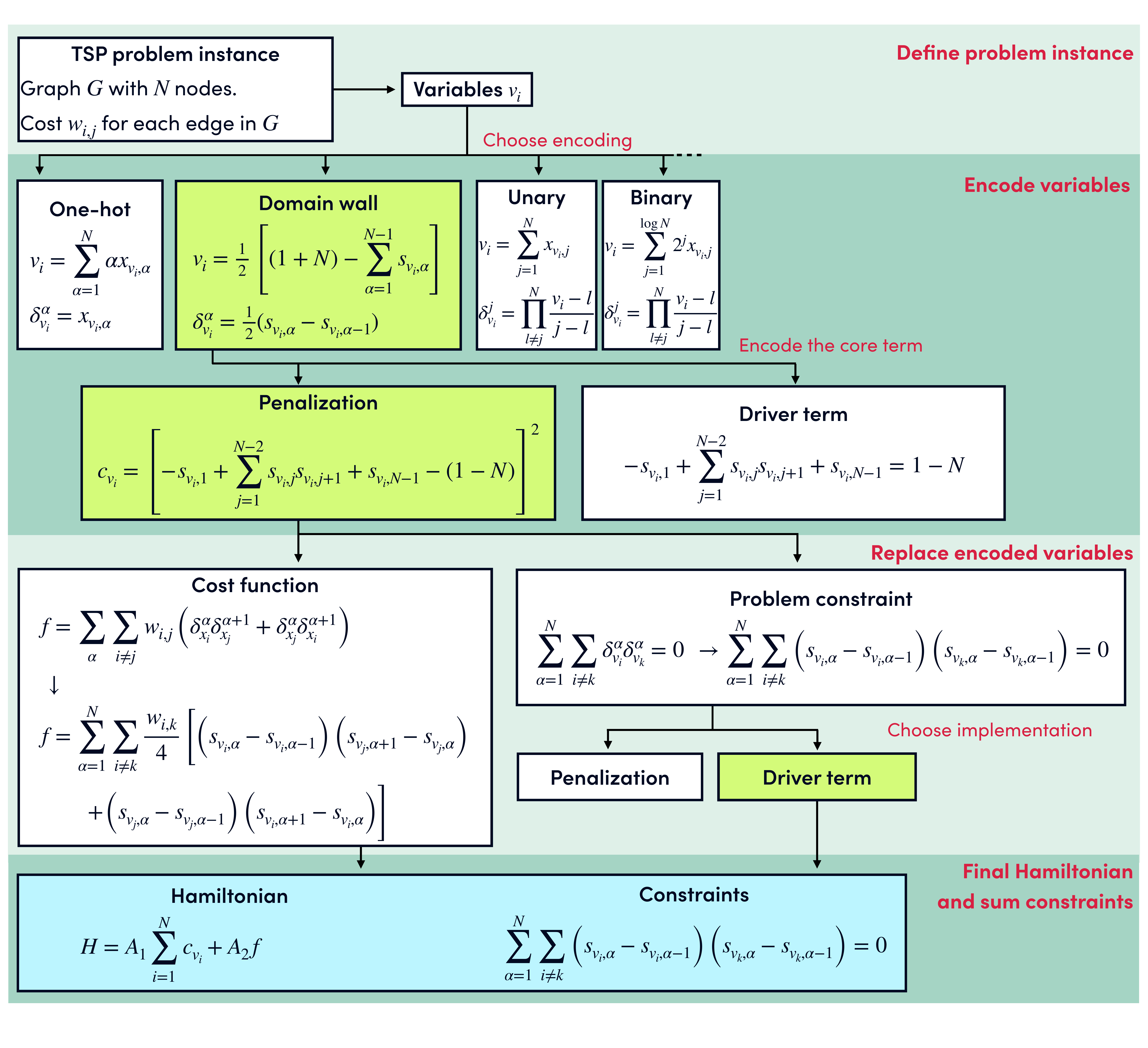}
              \caption{Example decision tree for the TSP problem. The Hamiltonian is obtained by substituting in the encoding-independent expressions of Eqs.~\eqref{eq:tsp_cost} and~\eqref{eq:permutation_constraint} the value indicators according to the chosen encoding. The choices made in the decision tree are highlighted in green. For this problem instance we assume the graph $G$ is complete (there is an edge connecting every pair of nodes) and we choose domain-wall encoding. This sparse encoding is convenient since the cost function and the problem constraint include products of value indicators $\delta_{v_i}^{\alpha}\delta_{v_k}^{\beta}$, which would imply a large number of interaction terms using a dense encoding. Sparse encodings require a core term for defining the valid states of the Hilbert space. In this case, we choose to encode this core term as an energy penalization in the Hamiltonian while the problem constraint is considered as a sum constraint. The final Hamiltonian and the sum constraints can be directly encoded on a Parity chip using the Parity Compiler, without the need to reduce the problem to its QUBO formulation.} 
              \label{fig:TSP_tree}
            \end{figure*}

            In Fig.~\ref{fig:TSP_tree}, we present an example of a decision tree for this problem. The cost function and the constraint include products of value indicators $\delta_{v_i}^{\alpha}\delta_{v_k}^{\beta}$ for which sparse encodings like domain wall are better suited.

	\section{Problem library}\label{sec:prob_lib}

		\subsection{Partitioning problems}
			The common goal in partitioning problems is to look for partitions of a set $U$ minimizing a cost function $f$. A partition $P$ of $U$ is a set $P \coloneqq\left\lbrace U_k\right\rbrace_{k\in K}$ of subsets $U_k \subset U$, such that $U = \bigcup_{k=1}^K U_{k}$ and $U_k \cap U_{k'} = 0$ if $k\neq k'$. Partitioning problems require a discrete variable $v_i$ for each element $u_i\in U$. The value of the variable indicates to which subset the element belongs, so $v_i$ can take $K$ different values. The values assigned to the subsets are arbitrary, therefore the value of the variable $v_i$ is usually not important in these cases, but only the value indicator $\delta_{v_i}^{\alpha}$ is needed, so sparse encodings may be convenient for these variables. 
		
			\subsubsection{Clustering problem} \label{subsub: clustering_prob}
	            \paragraph{Description} 
	                Let $U=\left\{u_i\right\}_{i=1}^N$ be a set of $N$ elements, characterized by weights $w_i \in \{1, ..., w_{\mathrm{max}}\}$  and distances $d_{i,j}$ between them. The clustering problem looks for a partition of the set $U$ into $K$ non-empty subsets $U_k$ that minimizes the distance between vertices in each subset. Partitions are subject to a weight restriction: for every subset $U_k$, the sum of the weights of the vertices in the subset must not exceed a given maximum weight $W_{\mathrm{max}}$.
	            \paragraph{Variables}
	                We define a variable $v_i=1,\dots,K$ for each element in $U$. We also require an auxiliary variable $y_j=0,1,\dots,W_{\mathrm{max}}$ per subset $U_k$, that indicates the total weight of the elements in $U_k$:
	                \begin{equation}
                        y_k = \sum_{\{i:u_i\in U_k\}} w_i. 
	                \end{equation}
	                
	            \paragraph{Constraints}
	                The weight restriction is an inequality constraint: 
        	        \begin{equation}
	                    y_k \leq W_{\mathrm{max}},
                    \end{equation}
                    which can be expressed as:
                    \begin{equation} \label{eq:clustering_constraint}
                        y_k = \sum_{i} w_i \delta_{v_i}^{k}.
                    \end{equation}
                If the encoding for the auxiliary variables makes it necessary (e.g. a binary encoding and $W_{\mathrm{max}}$ not a power of $2$), this constraint must be shifted as is described in Sec.~\ref{bb:ineq}.
	            \paragraph{Cost function}
	                The sum of the distances of the elements of a subset is:
	                \begin{equation}
	                    \begin{aligned}
	                        f_k = \sum_{\{i<j:u_i,u_j\in U_k\}} d_{i,j} = \sum_{i<j} d_{i,j} \delta_{v_i}^{k}\delta_{v_j}^{k},
	                    \end{aligned}
                    \end{equation}
		            and the cost function results:
		            \begin{equation} \label{eq:clustering_cost}
		                f = \sum_{k} f_k.
		            \end{equation}
		            
		      \paragraph{References} This problem is studied in Ref.~\cite{feld2019hybrid} as part of the Capacitated Vehicle Routing problem. 

			\subsubsection{Number partitioning}
				
				\paragraph{Description} 
					Given a set $U$ of $N$ real numbers $u_i$, we look for a partition of size $K$ such that the sum of the numbers in each subset is as homogeneous as possible. 
				\paragraph{Variables} 
					The problem requires $N$ variables ${v_i\in [1,K]}$, one per element $u_i$. The value of $v_i$ indicates to which subset $u_i$ belongs.
				
				\paragraph{Cost function} 
						The partial sums can be represented using the value indicators associated to the variables $v_j$:
					\begin{equation}
						p_j = \sum_{u_i \in U_j} u_i = \sum_{i=1}^{N} u_i \delta_{v_i}^j.
					\end{equation}	
					There are three common approaches for finding the optimal partition: maximizing the minimum partial sum, minimizing the largest partial sum or minimizing the difference between the maximum and the minimum partitions. The latter option can be formulated as
					\begin{equation}
						f = \sum_{i< j} \left(  p_i - p_j \right)^2.
					\end{equation}
			        In order to minimize the maximum partial sum (maximizing the minimum partial sum is done analogously) we introduce an auxiliary variable $l$ that can take values $1, ..., \sum_i u_i \equiv l_{\mathrm{max}}$. Depending on the problem instance the range of $l$ can be restricted further. The first term in the cost function 
                    \begin{equation}
                        f = l_{\mathrm{max}} \left( 1 - \prod_i \Theta(l-p_i) \right) + l
                    \end{equation}
                    then enforces that $l$ is as least as large as the maximum $p_i$ (see Sec.~\ref{bb:min_maximum}) and the second term minimizes $l$. The theta step function can be expressed in terms of the value indicator functions according to the building block Eq.~\eqref{eq:theta} where we either have to introduce auxiliary variables or express the value indicators directly according to the discussion in Sec.~\ref{bb:helper_fct}.
				\paragraph{Special cases} 
					For $K=2$, the cost function that minimizes the difference between the partial sums is 
					\begin{equation}
						\begin{aligned}
							f &= \left(  p_2 - p_1 \right)^2
							\\ & = \left[ \sum_{i=1}^N \left( \delta_{v_i}^1 -\delta_{v_i}^2 \right)u_i\right]^2.
						\end{aligned}	
					\end{equation}
				The only two possible outcomes for $\delta_{v_i}^1 -\delta_{v_i}^2$ are $\pm 1$, so these factors can be trivially encoded using spin variables $s_i = \pm1$, leading to
				\begin{equation}
				    f = \sum_{i=1}^N s_i u_i.
				\end{equation}

                \paragraph{References}
                The Hamiltonian formulation for $K=2$ can be found in \cite{lucas2014ising}.
					
			\subsubsection{Graph coloring}	
			    \paragraph{Description} 
			        In this problem, the nodes of a graph $G=(V,E)$ are divided into $K$ different subsets, each one representing a different color. We look for a partition in which two adjacent nodes are painted with different colors, and we can also look for a partition that minimizes the number of color used. 
			    \paragraph{Variables}
			        We define a variable $v_i=1\dots K$ for each node $i = 1 ,..., N$ in the graph. 
			    \paragraph{Constraints} 
			        We must penalize solutions for which two adjacent nodes are painted with the same color. The cost function of the graph partitioning problem presented in Eq.~\eqref{eq:graph_part_cost_funct} can be used for the constraint of graph coloring. In that case, $\left[1 - \delta(v_i-v_j)\right] A_{i,j}$ was used to count the number of adjacent nodes belonging to different subsets, now we use $\delta(v_i-v_j)A_{i,j}$ to indicate if nodes $i$ and $j$ are painted with the same color. Therefore the constraint is (see building block~\ref{bb:all_different})
			        \begin{equation}
			             c = \sum_{i<j} \delta(v_i-v_j) A_{i,j} = 0.
			        \end{equation}
			        
		        \paragraph{Cost function}
                The decision problem (``Is there a coloring that uses $K$ colors?'') can be answered by implementing the constraint as a Hamiltonian ${H = c}$ (note that ${c\geq 0}$). The existence of a solution with zero energy implies that a coloring with $K$ colors exists. 
                
		        Alternatively we can look for the coloring that uses the minimum number of colors. To check if a color $\alpha$ is used, we can use the following term (see building block~\ref{bb:alpha_is_used}):
		            \begin{equation}
		                u_{\alpha} = 1- \prod_{i=1}^{N} \left(1-\delta_{v_i}^{\alpha} \right),
		            \end{equation}
		            which is one if and only if color $\alpha$ is not used in the coloring, and 0 if at least one node is painted with color $\alpha$. The number of colors used in the coloring is the cost function of the problem:
		            \begin{equation}
		                f = \sum_{\alpha=1}^{K}  u_{\alpha}.
		            \end{equation}
		            This objective function can be very expensive to implement, since $u_{\alpha}$ includes products of $\delta_{v_i}^{\alpha}$ of order $N$, so a good estimation of the minimum number $K_{\mathrm{min}}$ would be useful to avoid using an unnecessarily large $K$. 

                    \paragraph{References}
                    The one-hot encoded version of this problem can be found in Ref.~\cite{lucas2014ising}.
		
			\subsubsection{Graph partitioning}		
				\paragraph{Description} 
		           Graph partitioning divides the nodes of a graph $G=(V,E)$ into $K$ different subsets, so that the number of edges connecting nodes in different subsets (cut edges) is minimized or maximized. 
				\paragraph{Variables} 
				    We employ one discrete variable ${v_k=1,\dots,K}$ per node in the graph, which indicates to which partition the node belongs. 
				\paragraph{Cost function} 
				    An edge connecting two nodes $i$ and $j$ is cut when $v_i\neq v_j$. It is convenient to use the symbol (see building block~\ref{bb:compare variables}):
				    \begin{equation} \label{eq:vi_equal_vj_indicator}
				        \delta(v_i-v_j) = \sum_{\alpha=1}^{K} \delta_{v_i}^{\alpha} \delta_{v_j}^{\alpha}=\begin{cases}1\; \mathrm{if}\, v_i=v_j
			        \\ 0\; \mathrm{if}\, v_i\neq v_j
			        \end{cases}, 
				    \end{equation}
                    which is equal to 1 when nodes $i$ and $j$ belong to the same partition ($v_i=v_j$) and zero when not ($v_i\neq v_j$). In this way, the term:
                    \begin{equation}
                        \left(1 - \delta(v_i-v_j)\right) A_{i,j}
                    \end{equation}
                    is equal to 1 if there is a cut edge connecting nodes $i$ and $j$, being $A_{i,j}$ the adjacency matrix of the graph. The cost function for minimizing the number of cut edges is obtained by summing over all the nodes:
                    \begin{equation} \label{eq:graph_part_cost_funct}
                        f = \sum_{i<j} \left[1 - \delta(v_i-v_j)\right] A_{i,j}, 
                    \end{equation}
                    or alternatively $-f$ to maximize the number of cut edges. 
                \paragraph{Constraints} 
                    A common constraint imposes that partitions have a specific size. The element counting building block \ref{bb:ele_cnt} is defined as
                    \begin{equation}
                        t_{\alpha}=\sum_{i=1}^{|V|} \delta_{v_i}^{\alpha}. 
                    \end{equation}
                    If we want the partition $\alpha$ to have $L$ elements, then
                    \begin{equation}
                        t_{\alpha} = L
                    \end{equation}
                    must hold. If we want two partitions $\alpha$ and $\beta$ to have the same size, then the constraint is
                    \begin{equation} \label{eq:graph_part_equal_size}
                        t_{\alpha} = t_{\beta},
                    \end{equation}
                    and all the partitions will have the same size imposing
                    \begin{equation}
                        \sum_{a<b} (t_a -t_b)^2 = 0
                    \end{equation}
                    which is only possible if $|V|/N$ is a natural number. 

                    \paragraph{References}
                    The Hamiltonian for $K=2$ can be found in \cite{lucas2014ising}. 
  
                \paragraph{Hypergraph partitioning} The problem formulation can be extended to hypergraphs. A hyperedge is cut when contains vertices from at least two different subsets. Given a hyperedge $e$ of $G$, the function
			    \begin{equation}
			        \mathrm{cut}(e)=\prod_{i, j \in e} \delta(v_i-v_j)
			    \end{equation}
			    is only equal to 1 if all the vertices included in $e$ belong to the same partition, and zero in any other case. The product over the vertices $i,j$ of the edge $e$ only needs to involve pairs such that every vertex appears at least once. The optimization objective of minimizing the cut hyperedges is implemented by the sum of penalties
			    \begin{equation}
			        \begin{aligned}
			        f &=  \sum_{e \in E} \left( 1-\mathrm{cut}(e) \right)
			            \\ &=  \sum_{e \in E} \left( 1-\prod_{i, j \in e} \sum_{a =1}^K \delta^a_i \delta^a_j \right).
			        \end{aligned}
			    \end{equation}
			    This objective function penalizes all possible cuts in the same way, regardless of the number of vertices cut or the number of partitions to which an edge belongs.

			\subsubsection{Clique cover}
			    \paragraph{Description}
			    Given a graph ${G = (V, E)}$, we are looking for the minimum number of colors $K$ for coloring all vertices such that the subsets $W_{\alpha}$ of vertices with the color ${\alpha}$ together with the edge set $E_{\alpha}$ restricted to edges between vertices in $W_{\alpha}$ form complete graphs. A subproblem is to decide if there is a clique cover using $K$ colors.
			    \paragraph{Variables}
			    For each vertex ${i = 1,\dots, N}$ we define variables ${v_i = 1, \dots, K}$ indicating the color that vertex is assigned to. If the number of colors is not given one has to start with some initial guess or some minimal value for $K$.
			    \paragraph{Constraints}
			    In this problem ${G_{\alpha}=(W_{\alpha},E_{\alpha})}$ has to be a complete graph, so the maximum number of edges in $E_{\alpha}$ must be present. Using the element counting building block, we calculate the number of vertices with color ${\alpha}$ (see bulding block~\ref{bb:ele_cnt}):
			    \begin{equation}
			        t_\alpha = \sum_{i=1}^{N} \delta^{\alpha}_i.
			    \end{equation}
			    If $G_{\alpha}$ is a complete graph, then the number of edges in $E_{\alpha}$ is $t_{\alpha}(t_{\alpha}-1)/2$
			    and thus the constraint reads
			    \begin{equation}
			        c = \sum_{{\alpha}=1}^K \left( \frac{t_{\alpha}(t_{\alpha}-1)}{2} - \sum_{(i,j) \in E} \delta^{\alpha}_i \delta^{\alpha}_j \right) =0.
			    \end{equation}
			    Note that we do not have to square the term as it can never be negative.
			    \paragraph{Cost function} 
			        The decision problem (``Is there a clique cover using $K$ colors?'') can be answered using the constraint $c$ as the Hamiltonian of the problem. For finding the minimum number of colors $K_{\mathrm{min}}$ for which a clique cover exist (the \textit{clique cover number}), we add a cost function for minimizing $K$. As in the graph coloring problem, we minimize the number of colors using
			        \begin{equation}
		                f = \sum_{\alpha} \left(1- u_{\alpha} \right),
		            \end{equation}
		            where 
		            \begin{equation}
		                u_{\alpha} = \prod_{i=1}^{N} \left(1-\delta_{v_i}^{\alpha} \right)
		            \end{equation}
                    indicates if the color $\alpha$ is used or not (see building block~\ref{bb:alpha_is_used}).

                    \paragraph{References}
                    The one-hot encoded Hamiltonian of the decision problem can be found in Ref.~\cite{lucas2014ising}.

		\subsection{Constrained subset problems}
		Given a set $U$, we look for a non-empty subset $U_0 \subseteq U$ that minimizes a cost function $f$ while satisfying a set of constraints $c_i$. In general, these problems require a binary variable $x_i$ per element in $U$ which indicates if the element $i$ is included or not in the subset $U_0$. Although binary variables are trivially encoded in single qubits, non-binary auxiliary variables may be necessary to formulate constraints, so the encoding-independent formulation of these problems is still useful. 
		
		\subsubsection{Cliques}
		    \paragraph{Description}
		    A clique on a given graph $G = (V, E)$ is a subset of vertices $W \subseteq V$ such that $W$ and the subset $E_W$ of edges between vertices in $W$ is a complete graph, i.e. the maximal possible number of edges in $E_W$ is present. The goal is to find a clique with cardinality $K$. Additionally one could ask what the largest clique of the graph is.
		    \paragraph{Variables}
		    We define $|V|$ binary variables $x_i$ that indicate whether vertex $i$ is in the clique or not.
		    \paragraph{Constraints}
		    This problem has two constraints, namely that the cardinality of the clique is $K$ and that the clique indeed has the maximum number of edges.
		    The former is enforced by
		    \begin{equation}
		        c_1 = \sum_{i=1}^{|V|} x_i = K 
		    \end{equation}
		    and the latter by
		    \begin{equation}
		        c_2 = \sum_{(i,j) \in E} x_i x_j=\frac{K(K-1)}{2}  .
		    \end{equation}
		    If constraints are implemented as energy penalization, one has to make sure that the first constraint is not violated in order to decrease the penalty for the second constraint. Using the cost/gain analysis (see Sec.~\ref{bb:prio}) of a single spin flip this is prevented as long as $a_1 \gtrapprox  a_2  \Delta$, where $\Delta$ is the maximal degree of $G$ and $a_{1,2}$ are the energy scales of the first and second constraints.
		    
		    \paragraph{Cost function}
		    The decision problem (is there a clique of size $K$?) can be solved using the constraints as the Hamiltonian of the problem. If we want to find the largest clique of the graph $G$, we must encode $K$ as a discrete variable, $K=1,\dots,K_{\mathrm{max}}$, where $K_{\mathrm{max}}= \Delta$ is the maximum degree of $G$ and the largest possible size of a clique. The cost function for this case is simply the value of $K$:
		    \begin{equation}
		        f = -K 
		    \end{equation}
		    \paragraph{Resources}
		    Implementing constraints as energy penalizations, the total cost function for the decision problem (fixed $K$), 
		    \begin{equation}
		        f = \left(c_1-K \right)^2 + \left(c_2-\tfrac{K(K-1)}{2} \right)^2,
		    \end{equation}
		    has maximum order two of the interaction terms and the number of terms scales with $|E| + |V|^2$. If $K$ is encoded as a discrete variable, the resources depend on the chosen encoding. 

            \paragraph{References}
            This Hamiltonian was formulated in Ref.~\cite{lucas2014ising} using one-hot encoding.

		\subsubsection{Maximal independent set}
		\paragraph{Description} Given an hypergraph $G=(V,E)$ we look for a subset of vertices $S\subset V$ such that there are no edges in $E$ connecting any two vertices of $S$. Finding the largest possible $S$ is an NP-hard problem. 
		
		\paragraph{Variables} We use a binary variable $x_i$ for each vertex in $V$.
		
		\paragraph{Cost function} For maximizing the number of vertices in $S$, the cost function is
		\begin{equation}
		    f = -\sum_{i=1}^{|V|} x_i.
		\end{equation}
		
		\paragraph{Constraints} Given two elements in $S$, there must not be any edge of hyperedge in $E$ connecting them. The constraint
		\begin{equation}
		    c = \sum_{i,j \in V} A_{i,j} x_ix_j 
		\end{equation}
		counts the number of adjacent vertices in $S$, being $A$ the adjacency matrix. Setting $c=0$, the vertices in $S$ form an independent set.  

        \paragraph{References}
        This problem can be found in Refs.~\cite{lucas2014ising, choi2010adiabatic} for graphs.

		\subsubsection{Set packing}
		    \paragraph{Description}
		    Given a set $U$ and a family $S = \{V_i\}_{i= 1}^{N} $ of subsets $V_i$
		    of $U$, we want to find set packings, i.e., subsets of $S$ such that all subsets are pairwise disjoint, $V_i \cap V_j = \emptyset$. Finding the maximum packing (the maximum number of subsets $V_i$) is the NP-hard optimization problem called set packing.
		    \paragraph{Variables}
		    We define $N$ binary variables $x_i$ that indicate whether subset $V_i$ belongs to the packing.
		    \paragraph{Cost function}
		    Maximizing the number of subsets in the packing is achieved with the element counting building block
		    \begin{equation}
		        f = - \sum_{i=1}^N x_i. 
		    \end{equation}
		    \paragraph{Constraints}
		    In order to ensure that any two subsets of the packings are disjoint we impose a cost on overlapping sets with
		    \begin{equation}
		        c =  \sum_{i, j: V_i \cap V_j \neq \emptyset} x_i x_j = 0.
		    \end{equation}
		    
		    \paragraph{Resources}
		    The total cost function $H = H_A + H_B$ has maximum order two of the interaction terms (so it is a QUBO problem) and the number of terms scales with up to $N^2$.

            \paragraph{References}
            This problem can be found in Refs.~\cite{lucas2014ising}.
		
		\subsubsection{Vertex cover}
		    \paragraph{Description}
		    Given a hypergraph ${G = (V, E)}$ we want to find the smallest subset ${C \subseteq V}$ such that all edges contain at least one vertex in $C$.
		    \paragraph{Variables}
		    We define $|V|$ binary variables $x_i$ that indicate whether vertex $i$ belongs to the cover $C$.
		    \paragraph{Cost function}
		    Minimizing the number of vertices in $C$ is achieved with the element counting building block
		    \begin{equation}
		        f =  \sum_{i =1}^{|V|} x_i.
		    \end{equation}
		    \paragraph{Constraints}
		    With 
		    \begin{equation}
		        c= \sum_{e = (u_1,..., u_k) \in E}\,  \prod_{a=1}^k(1-x_{u_a})=0
		    \end{equation}
		    one can penalize all edges that do not contain vertices belonging to $C$. Encoding the constraint as an energy penalization, the Hamiltonian results in
		    \begin{equation}
		        H = Af+Bc.
		    \end{equation}
		    By setting $B > A$ we avoid that the constraint is traded off against the minimization of $C$.
		    \paragraph{Resources}
		    The maximum order of the interaction terms is the maximum rank of the hyperedges $k$ and the number of terms scales with $|E| 2^k+ |V|$.

            \paragraph{References}
            The special case that only considers graphs can be found in Ref.~\cite{lucas2014ising}.
		
		\subsubsection{Minimal maximal matching}
		    \paragraph{Description}
		    Given a hypergraph $G = (V, E)$ with edges of maximal rank $k$ we want to find a minimal (i.e. fewest edges) matching $C \subseteq E$ which is maximal in the sense that all edges with vertices which are not incident to edges in $C$ have to be included to the matching.
		    \paragraph{Variables}
		    We define $|E|$ binary variables $x_i$ that indicate whether an edge belongs to the matching $C$.
		    \paragraph{Cost function}
		    Minimizing the number of edges in the matching is simply done by the cost function
            \begin{equation}
                f_A =A \sum_{i=1}^{|E|} x_i.
            \end{equation}
            
            \paragraph{Constraints}
            We have to enforce that $C$ is indeed a matching, i.e. that no two edges which share a vertex belong to $C$. Using an energy penalty this is achieved by:
            \begin{equation}
                f_B = B \sum_{v \in V} \sum_{(i, j) \in \partial v} x_i x_j=0,
            \end{equation}
		    where $\partial v$ is the set of edges connected to vertex $v$. Additionally, the matching should be maximal. For each vertex $u$, we define a variable $y_u = \sum_{i \in \partial u} x_i$ which is only zero if the vertex does not belong to an edge of $C$. If the first constraint is satisfied, this variable can only be $0$ or $1$. In this case, the constraint can be enforced by
		    \begin{equation}
		        f_C = C \sum_{e =(u_1, ..., u_k) \in E} \prod_{a=1}^k(1-y_{u_a})=0.
		    \end{equation}
		    However, one has to make sure that the constraint implemented by $f_B$ is not violated in favor of $f_C$ which could happen if for some $v$, $y_v > 1$ and for $m$ neighboring vertices $y_u = 0$. Then the contributions from $v$ are given by
		    \begin{equation}
		        f_v = B y_v (y_v -1) \frac{1}{2} + C (1-y_v) m
		    \end{equation}
		    and since $m + y_v$ is bounded by the maximum degree $\Delta$ of $G$ times the maximum rank of the hyperedges $k$, we need to set ${B > (\Delta k-2)C}$ to ensure that the ground state of ${f_B + f_C}$ does not violate the first constraint.
		    Finally, one has to prevent $f_C$ to be violated in favor of $f_A$ which entails $C > A$.
		    \paragraph{Resources}
		    The maximum order of the interaction terms is $k$ and the number of terms scales roughly with $(|V|+ |E|2^k) \Delta (\Delta-1)$.

            \paragraph{References}
            This problem can be found in Ref.~\cite{lucas2014ising}.
      
		\subsubsection{Set cover}
			\paragraph{Description}
				Given a set ${U=\lbrace u_{\alpha} \rbrace_{\alpha=1}^{n}}$ and $N$ subsets ${V_i \subseteq U}$, we look for the minimum number of $V_i$ such that ${U = \bigcup V_i}$.
			\paragraph{Variables}
				We define a binary variable ${x_i=0,1}$ for each subset $V_i$ that indicates if the subset $V_i$ is selected or not. We also define auxiliary variables $y_{\alpha}=1,2,\dots,N$ that indicate how many active subsets ($V_i$ such that ${x_i=1}$) contain the element $u_{\alpha}$. 
			\paragraph{Cost function}
				The cost function is simply
				\begin{equation}
					f = \sum_{i=1}^{N}x_i,
				\end{equation}
				which counts the number of selected subsets $V_i$. 
			\paragraph{Constraints}
				The constraint  $U = \bigcup_{i:x_i=1} V_i$ can be expressed as 
				\begin{equation}
					y_{\alpha}> 0, \; \forall \alpha=1,\dots,n,
				\end{equation}
				 which implies that every element $u_{\alpha}\in U$ is included at least once. These inequalities are satisfied if $y_{\alpha}$ are restricted to the valid values ($y_{\alpha}=1,2,\dots,N$, $y_{\alpha}\neq 0$) (see Sec.~\ref{bb:ineq}). The values of $y_{\alpha}$ should be consistent with those of $x_i$, so the constraint is
				\begin{equation}
					c_{\alpha}=  y_{\alpha} - \sum_{i: u_{\alpha\in V_i}}x_i =0.
				\end{equation}	
				
			\paragraph{Special case: exact cover} If we want each element of $U$ to appear once and only once on the cover, then $y_{\alpha} =1, $ for all $ \alpha$ and the constraint of the problem reduces to
				\begin{equation}
				    c_{\alpha} = \sum_{i: \alpha\in V_i}x_i =1.
				\end{equation}

            \paragraph{References}
            The one-hot encoded Hamiltonian can be found in Ref.~\cite{lucas2014ising}.

		\subsubsection{Knapsack}
			\paragraph{Description}
				A set $U$ contains $N$ objects, each of them with a value $d_i$ and a weight $w_i$. We look for a subset of $U$ with the maximum value $\sum d_i$ such that the total weight of the selected objects does not exceed the upper limit $W$. 
			\paragraph{Variables}
				We define a binary variable $x_i=0,1$ for each element in $U$ that indicates if the element $i$ is selected or not. We also define an auxiliary variable $y$ that indicates the total weight of the selected objects:
				\begin{equation}
					y = \sum_{i:x_i=1}w_i.
				\end{equation}
			If the weights are natural numbers $w_i \in \mathbb{N}$ then $y$ is also natural, and the encoding of this auxiliary variable is greatly simplified. 
			\paragraph{Cost function}
				The cost function is given by
				\begin{equation}
					f =-\sum_{i=1}^{N} d_i x_i,
				\end{equation}
				which counts the value of the selected elements. 
			\paragraph{Constraints}
				The constraint  ${y<W}$ is implemented by forcing $y$ to take one of the possible values $y=1,\dots W-1$ (see Ref.~\ref{bb:ineq}). The value of $y$ must be consistent with the selected items from $U$: 
				\begin{equation}
					c =  y - \sum_{i}^N w_i x_i =0.
				\end{equation}					
            \paragraph{References}
            The one-hot encoded Hamiltonian can be found in Ref.~\cite{lucas2014ising}.
		
		\subsection{Permutation problems}
			In permutation problems, we need to find a permutation of $N$ elements that minimizes a cost function while satisfying a given set of constraints. In general, we will use a discrete variable ${v_i\in [1,N]}$ that indicates the position of the element $i$ in the permutation. 
				\subsubsection{Hamiltonian cycles} 
					\paragraph{Description} 
						For a graph ${G=(V,E)}$,  we ask if a Hamiltonian cycle, i.e., a closed path that connects all nodes in the graph through the existing edges without visiting the same node twice, exists.
					\paragraph{Variables} 
						We define a variable ${v_i =1,\dots,|V|}$ for each node in the graph, that indicates the position of the node in the permutation. 
					\paragraph{Cost function} 
						For this problem there is no cost function, so every permutation that satisfies the constraints is a solution of the problem. 
					\paragraph{Constraints}\label{para:HamcycleConstraints}
						This problem requires two constraints. The first constraint is inherent to all permutation problems, and imposes to the $|V|$ variables $\lbrace v_i \rbrace$ to be a permutation of $[1,\dots,|V|]$. This is equivalent to require $v_i\neq v_j$ if $i \neq j$, which can be encoded with the following constraint (see Sec.~\ref{bb:all_different}):
						\begin{equation}\label{eq:permutation_constraint}
							c_1 =  \sum_{\alpha=1}^{N}\sum_{i\neq k} \delta_{v_i}^{\alpha}\delta_{v_k}^{\alpha} = 0 .
						\end{equation}
						The second constraint ensures that the path only goes through the edges of the graph. Let $A$ be the adjacency matrix of the graph, such that $A_{i,j}=1$ if there is an edge connecting nodes $i$ and $j$ and zero otherwise. To penalize invalid solutions, we use the constraint
						\begin{equation}
							c_2 = \sum_{i,j }  \sum_{\alpha=1}^{|V|} (1-A_{i,j})\left(\delta_{v_i}^{\alpha}\delta_{v_j}^{\alpha+1} + \delta_{v_i}^{\alpha+1}\delta_{v_j}^{\alpha}  \right)=0,
						\end{equation} 
						which counts how many adjacent nodes in the solution are not connected by an edge in the graph. $\alpha = |V|+1$ represents $\alpha=1$ since we are looking for a closed path. 

                    \paragraph{References}
                    The one-hot encoded Hamiltonian can be found in Ref.~\cite{lucas2014ising}.

				\subsubsection{Traveling salesperson problem (TSP)} \label{sec:TSP_problem}
					\paragraph{Description} 
						The TSP is a trivial extension of the Hamiltonian cycles problem. In this case, the nodes represent cities and the edges are the possible roads connecting the cities, although in general it is assumed that all cities are connected (the graph is complete). For each edge connecting cities $i$ and $j$ there is a cost $w_{i,j}$. The solution of the TSP is the Hamiltonian cycle that minimizes the total cost $\sum{w_{i,j}}$.
					\paragraph{Variables} 
						We define a variable ${v_i =1,\dots,|V|}$ for each node in the graph, indicating the position of the node in the permutation. 
					\paragraph{Cost function}
					    If the traveler goes from city $i$ at position $\alpha$ to city $j$ in the next step, then 
					    \begin{equation}
					        \delta_{v_i}^{\alpha}\delta_{v_j}^{\alpha+1} = 1,    
					    \end{equation}
                        otherwise that expression would be zero.
					    Therefore the total cost of the travel is codified in the function:
						\begin{equation} \label{eq:tsp_cost}
							f = \sum_{\alpha} \sum_{i<j} w_{i,j} \left( \delta_{v_i}^{\alpha}\delta_{v_j}^{\alpha+1} +  \delta_{v_j}^{\alpha}\delta_{v_i}^{\alpha+1}\right).
					\end{equation} 
					\paragraph{Constraints} 
						The constraints are the same as those used in the Hamiltonian Cycles problem (see paragraph ~\ref{para:HamcycleConstraints}). If the graph is complete (all the cities are connected) then constraint $c_2$ is not necessary. 

                    \paragraph{References}
                    The one-hot encoded Hamiltonian can be found in Ref.~\cite{lucas2014ising}.
						
				\subsubsection{Machine scheduling}\label{sec:machine-scheduling}
					\paragraph{Description} Machine scheduling problems seek the best way to distribute a number of jobs over a finite number of machines. Basically, these problems explore permutations of the job list, where the position of a job in the permutation indicates on which machine and at what time the job is executed. Many variants of the problem can be found in the literature, including formulations of the problem in terms of spin Hamiltonians~\cite{kurowski2020hybrid,venturelli2015quantum,amaro2022case}. Here we consider the problem of $M$ machines and $N$ jobs, where all jobs take the same amount of time to complete, so the time can be divided into time slots of equal duration $t$. It is possible to include jobs of duration $nt$ ($n\in \mathbb{N}$) by using appropriate constraints that force some jobs to run in consecutive time slots on the same machine. In this way, problems with jobs of different duration can be solved by choosing the time $t$ sufficiently small.
					
					\paragraph{Variables}   We define variables $v_{m,t}=0,\dots N$, where the subindex $m=1,\dots M$ indicates the machine and $t=1,\dots T$ the time slot. When $v_{m,t}=0$, the machine $m$ is unoccupied in time slot $t$, and if $v_{m,t}=j\neq 0$  then the job $j$ is done in the machine $m$, in the time slot $t$. 
					
					\paragraph{Constraints} There are many possible constraints depending on the use case we want to run. As in every permutation problem, we require that no pair of variables have the same value, $v_{m,t}\neq v_{m',t'}$, otherwise some jobs would be performed twice. We also require each job to be complete so there must be exactly one $v_{m,t}=j$ for each job $j$. This constraint is explained in Sec.~\ref{bb:all_different} and holds for every job $j\neq 0$: 
					
					\begin{equation}
						c_{1} =  \sum_{j\neq0}\left[\sum_{m,t}  \delta_{v_{m,t}}^{j}-1\right]^2 =0.
					\end{equation}
					Note that if some job $j$ is not assigned (i.e. there are no $m,t$ such that $v_{m,t}=j$) then $c_1>0$. Also if there is more that one variable $v_{m,t}=j$, then again $c_1>0$. The constraint will be satisfied ($c_1=0$) if and only if every job is assigned to a single time slot, on a single machine. 
					
					Suppose job $k$ can only be started if another job $j$ has been done previously. This constraint can be implemented as
					\begin{equation}
						c_{2,k>j} = \sum_{m,m',t\geq t'} \delta_{v_{m,t}}^{j} \delta_{v_{m',t'}}^{k}=0,
					\end{equation}
    				which precludes any solution in which job $j$ is done after job $k$. Alternatively, it can be codified as
    				\begin{equation}
						c^{\prime}_{2,k>j} = \sum_{m,m',t< t'} \delta_{v_{m,t}}^{j} \delta_{v_{m',t'}}^{k}=1.
					\end{equation}
    				
    				These constraints can be used to encode problems that consider jobs with different operations $O_1,\dots O_q$ that must be performed in sequential order. Note that constraints $c_2,c^{\prime}_2$ allow using different machines for different operations. If we want job $k$ to be done immediately after work $j$, we can substitute $t'$ by $t+1$ in $c^{\prime}_2$.
    				
    				If we want two jobs $j_1$ and $j_2$ to run in the same machine in consecutive time slots, the constraint can be encoded as
    				\begin{equation}
    					c_3 = \sum_{m, t} \delta_{v_{m,t}}^{j_1} \delta_{v_{m,t+1}}^{j_2}=1,
    				\end{equation}
    				or as a reward term in the Hamiltonian,
    				\begin{equation}
    					c_3^{\prime} = -\sum_{m, t} \delta_{v_{m,t}}^{j_1} \delta_{v_{m,t+1}}^{j_2},
    				\end{equation}
    				that reduces the energy of any solution in which job $j_2$ is performed immediately after job $j_1$ in the same machine $m$. This constraint allows to encode problems with different job duration, since $j_1$ and $j_2$ can be considered part of a same job of duration $2t$.
    				
    				\paragraph{Cost function} 
    				
    				Different objective functions can be chosen for this problem, such as minimizing machine idle time or early and late deliveries. A common option is to minimize the makespan, i.e., the time slot of the last scheduled job. To do this we first introduce an auxiliary function $\tau\left( v_{m,t}\right)$ that indicates the time slots in which some job has been scheduled:
    				\begin{equation}
    				    \tau\left( v_{m,t}\right) = \begin{cases}
    				    t \:\:\: &\mathrm{if}\;v_{m,t} \neq 0
    				    \\ 0 \:\:\: &\mathrm{if}\;v_{m,t} = 0.
				                                \end{cases}
    				\end{equation}
    				These functions can be generated from $v_{m,t}$:
    				\begin{equation}
    				    \tau\left( v_{m,t}\right) = \left(1-\delta_{v_{m,t}}^0 \right).
    				\end{equation}
    				Note that the maximum value of $\tau$ corresponds to the makespan of the problem, which is to be minimized. To do this, we introduce the extra variable $\tau_{\mathrm{max}}$ and penalize configurations where $\tau_{\mathrm{max}}<\tau(v_{m,t})$ for all $\; m,t$:
    				\begin{equation}
    				    f_{\tau_{\mathrm{max}}} = \prod_{m,t} \Theta\left[\tau(v_{m,t})-\tau_{\mathrm{max}}\right],
    				\end{equation}
    				with
    				\begin{equation}
    				    \Theta(x)=\begin{cases}1 \qquad \mathrm{if}\;x\geq 0
    				    \\0 \qquad \mathrm{if}\;x< 0.
    				    \end{cases}
    				\end{equation}
    				For details on how $\Theta(x)$ can be expressed see Sec.~\ref{bb:step_func}. 
    				By minimizing $f_{\tau_{\mathrm{max}}}$, we ensure that $\tau_{\mathrm{max}}\geq \tau(v_{m,t})$. 
    				Then the cost function is simply to minimize $\tau_{\mathrm{max}}$, the latest time a job is scheduled, via
    				\begin{equation}
    				    f = \tau_{\mathrm{max}}.
    				\end{equation}
    				
    				An alternative cost function for this problem is
    				\begin{equation}
    				    f^{\prime} = \sum_{m,t}\tau(v_{m,t}),
    				\end{equation}
				    which forces all jobs to be schedule as early as possible. 

                    \paragraph{References}
                    Quantum formulations of this and related problems can be found in Refs.~\cite{venturelli2015quantum,kurowski2020hybrid,amaro2022case,carugno2022evaluating}
        
				\subsubsection{Nurse Scheduling problem}\label{sec:nurse_scheduling}
				    \paragraph{Description} In this problem we have $N$ nurses and $D$ working shifts. Nurses must be scheduled with minimal workload following hard and soft constraints, like minimum workload $n_{\mathrm{min}, t}$ of nurses $i$ (where nurse $i$ contributes workload $p_i$) in a given shift $t$ and balancing the number of shifts as equal as possible. Furthermore, no nurse should have to work on more than $d_{\mathrm{max}}$ consecutive days.
				    \paragraph{Variables} 
				We define $N D$ binary variables $v_{i, t}$ indicating whether nurse $i$ is scheduled for shift $t$.
				
				\paragraph{Cost function}
				The cost function whose minimum corresponds to minimal number of overall shifts is given by
				\begin{equation}
				    f = \sum_{t=1}^D \sum_{i=1}^N v_{i,t}
				\end{equation}
				
				\paragraph{Constraints}
				Balancing of the shifts is expressed by the constraint
                \begin{equation}
                    c = \sum_{i<j} \left(\sum_t v_{i, t} - \sum_{t'} v_{j, t'} \right)^2.
                \end{equation}
                In order to get the minimal workload per shift we 
                introduce the auxiliary variables $y_t$ which we bind to the values
                \begin{equation}\label{eq:ns-aux-var}
                    y_t = \sum_{i=1}^N p_i v_{i,t}
                \end{equation}
                 with the penalty terms
                \begin{equation}\label{eq:ns-bind}
                    \left(y_t - \sum_{i=1}^N p_i v_{i,t}\right)^2.
                \end{equation}
                Now the constraint takes the form
                \begin{equation}
                    c= \sum_{t=1}^D \sum_{\alpha < n_{\mathrm{min}, t}} \delta^{\alpha}_{y_t} = 0. 
                \end{equation}
                Note that we can also combine the cost function for minimizing the number of shifts and this constraint by using the cost function
                \begin{equation}\label{eq:ns-constraint}
                    f =\sum_{t=1}^D \left( \sum_{i=1}^N p_i v_{i,t} - n_{\mathrm{min}, t} \right)^2.
                \end{equation}
                Finally, the constraint that a nurse $i$ should work in maximally $d_{\mathrm{max}}$ consecutive shifts reads
                \begin{equation}
                    c(i) = \sum_{t_1-t_2 > d_{\mathrm{max}}} \prod_{t'=t_1}^{t_2} v_{i,t'} = 0.
                \end{equation}

                \paragraph{References}
                This problem can be found in Ref.~\cite{ikeda2019application}.

		\subsection{Real variables problems}\label{sec:real_vars}

			Some problems are defined by a set of real variables $\{v_i \in \mathbb{R}, i = 1, ..., d \}, $ and a cost function $f(v_i)$ that has to be minimized. Just as for the discrete problems we discussed previously, we want to express these problems as Hamiltonians whose ground state represents an optimal solution. This shall be done in three steps; first we will encode the continuous variables to discrete ones. Second, we have to choose an encoding of these discrete variables to spin variables exactly as before and finally a decoding step is necessary that maps the solution to our discrete problem back to the continuous domain. In the following we investigate two possible methods one could use for the first step.
			
		    \paragraph{Standard discretization}
		    Given a vector $v \in \mathbb{R}^d$ one could simply discretize it by partitioning the axis into intervals with the length of the aspired precision $q$, i.e., the components of ${v = (v_i) , i = 1, ..., d}$ are mapped to ${v_i \rightarrow \tilde v_i \in \mathbb{N} }$ such that ${(\tilde{v}_i-1) q \leq v_i < \tilde v_i q}$. 
		    
		    Depending on the problem it might also be useful to have different resolutions for different axis or a non-uniform discretization, e.g., logarithmic scaling of the interval length. 
		    
		    For the discrete variables $\tilde v_i$ one can then use the encodings in Sec.~\ref{sec:enc_lib}. 
		    In the final step after an intermediate solution $\tilde v^*$ is found one has to map it back to $\mathbb{R}^d$ by uniformly sampling the components of $v^*$ from the hypercuboids corresponding to $\tilde v^*$.
            Any intermediate solution that is valid for the discrete encoding can also be decoded and thus there are no core terms in the cost function besides those from the discrete encoding.
            
            \paragraph{Random subspace coding}
            A further possibility to encode a vector $v \in \mathbb{R}^d$ into discrete variables  is random subspace coding~\cite{rachkovskii2005properties}. One starts with randomly choosing a set of coordinates ${D_n\coloneqq\left\lbrace 1,2,\dots, d_n\right\rbrace\subset \left\lbrace 1,2,\dots, d\right\rbrace}$. For each $i\in D_n$, a Dirichlet process is used to pick an interval $[a_i,b_i]$ in the $i$-coordinate direction in $\mathbb{R}^d$ \cite{devroye1993generating}. We denote
            \begin{equation}
            \begin{aligned}
                \pi_i:\mathbb{R}^d&\rightarrow \mathbb{R}\\
                x=(x_1,\dots, x_d)&\mapsto \pi_i(x)\coloneqq x_i
            \end{aligned}
            \end{equation}
            for the projection to the $i$-th coordinate. A \textit{hyperrectangle} $R\subset \mathbb{R}^d $ is defined as 
            \begin{equation}
                R\coloneqq \left\lbrace x\in \mathbb{R}^d\, \middle|\,  \pi_i(x)\in [a_i,b_i],\, \forall \, i\in D_n\right\rbrace.
            \end{equation}
            
            For the fixed set of chosen coordinates $D_n$ the Dirichlet processes are run $m$ times to get $m$ hyperrectangles $\left\lbrace R_1,\dots, R_m\right\rbrace $. For $k=1,\dots, m$, we define binary projection maps
            \begin{equation}
                \begin{aligned}
                    z_k:\mathbb{R}^d&\rightarrow \mathbb{Z}_2\\
                    x&\mapsto z_k(x)\coloneqq \begin{cases}
                    1,\, &\text{if}\, x\in R_k\\
                    0,\, &\text{else}.
                    \end{cases}
                \end{aligned}
            \end{equation}
            Then \textit{Random subspace coding} is defined as a map
            \begin{equation}
                \begin{aligned}
                    z:\mathbb{R}^d&\rightarrow \mathbb{Z}_2^m\\
                    x&\mapsto z(x)\coloneqq (z_1(x),\dots, z_m(x))\, .
                \end{aligned}
            \end{equation}
            
            Depending on the set of hyperrectangles $\left\lbrace R_1,\cdots, R_n\right\rbrace $, random subspace coding can be a sparse encoding. Let $M\coloneqq \left\lbrace 1, \dots, m\right\rbrace$ and $K\subset M$. We define sets 
            \begin{equation}
                \begin{aligned}
                   U(K)\coloneqq \bigcap_{k\in K}R_k- \bigcup_{i\in M-K} R_i \neq \emptyset\, . 
                \end{aligned}
            \end{equation}
            For $z\in \mathbb{Z}_2^m$, let $K(z)=\left\lbrace k\in \left\lbrace 1,\dots, m\right\rbrace \, \middle|\, z_k=1\right\rbrace$. The binary vector $z$ is in the image of the random subspace encoding if and only if 
            \begin{equation}
                \begin{aligned}
                    U(K(z))\neq \emptyset
                \end{aligned}
            \end{equation}
            holds. From here one can simply use the discrete encodings discussed in Sec.~\ref{sec:enc_lib} to map the components $z_i(x)$ of $z(x)$ to spin variables.
            Due to the potential sparse encoding, a core term has to be added to the cost function. Let $N=\left\lbrace L\subset M\, \middle|\, U(L)=\emptyset\right\rbrace$, then the core term reads
		    \begin{equation}\label{def: rsc_core}
		        c = \sum_{L\in N} \prod_{\ell\in L}\delta_{z_{\ell}}^1\prod_{p\in M-L}\delta_{z_{p}}^0\, . 
		    \end{equation}
		    Despite this drawback, random subspace coding might be preferable due to its simplicity and high resolution with relatively few hyperrectangles compared to the hypercubes of the standard discretization.
		    
			\subsubsection{Financial crash problem}
            
            \paragraph{Description}
            We are going to calculate the financial equilibrium of market values $v_i, i=1, ..., n$ of $n$ institutions according to a simple model following Refs.~\cite{elliott2014financial, orus2019forecasting}. In this model, the prices of $m$ assets are labeled by $p_k, \; k=1, ..., m$. Furthermore we define the ownership matrix $D$, where $D_{ij}$ denotes the percentage of asset $j$ owned by $i$, the cross-holdings $C$, where $C_{ij}$ denotes the percentage of institution $j$ owned by $i$ (except for self-holdings) and the self-ownership matrix $\tilde C$. The model postulates that without crashes the equity values $V$ (such that $v = \tilde C V$) in equilibrium satisfy
            \begin{equation}
                V = Dp + CV \rightarrow v = \tilde C (1-C)^{-1} D p.
            \end{equation}
            Crashes are then modeled as abrupt changes in the prices of assets held by an institution, i.e., via 
            \begin{equation}\label{eq:fin_eq}
                v = \tilde C (1-C)^{-1}( D p - b(v, p)),
            \end{equation}
            where $b_i(v, p) = \beta_i(p) (1- \Theta(v_i -v_i^c))$ results in the problem being highly non-linear.
            \paragraph{Variables}
            It is useful to shift the market values and we end up with a variable $v - v^c = v' \in \mathbb{R}^n$. Since the crash functions $b_i(v, p)$ explicitly depend on the components of $v'$ it is not convenient to use the random subspace coding (or only for the components individually) and so we employ the standard discretization, i.e. each component of $v'$ takes discrete values $0, ..., K$ such that with desired resolution $r$ a cut-off value $rK$ is reached.
            \paragraph{Cost function}
            In order to enforce that the system is in financial equilibrium we simply square Eq.~\eqref{eq:fin_eq}
            \begin{equation}
                f = \left( v' + v^c - \tilde C (1-C)^{-1}( D p -b(v,p) ) \right)^2,
            \end{equation}
            where for the theta functions in $b(v, p)$ we use
            \begin{equation}
                \Theta(v'_i) = \sum_{\alpha \geq 0} \delta^{\alpha}_{v'_i},
            \end{equation}
            as suggested in Sec.~\ref{bb:compare variables}.
            
            \subsubsection{Continuous black box optimization}
            
            \paragraph{Description}
            Given a function ${f: [0, 1]^d \rightarrow \mathbb{R}}$ which can be evaluated at individual points a finite number of times but where no closed form is available we want to find the global minimum. In classical optimization the general strategy is to use machine learning to learn an analytical acquisition function $g(x)$ in closed form from some sample evaluations, optimize it to generate the next point where to evaluate $f$ on and repeat as long as resources are available. 
            \paragraph{Variables}
            The number and range of variables depends on the continuous to discrete encoding. In the case of the standard discretization we have $d$ variables $v_a$ taking values in $\{0, ..., \lceil 1/q \rceil\}$, where $q$ is the precision. 
            
            With the random subspace encoding we have $\Delta$ variables $v_a$ taking values in $\{0, ..., d_s \}$, where $\Delta$ is the maximal overlap of the rectangles and $d_s$ is the number of rectangles.
            \paragraph{Cost function}
            Similar to the classical strategy we first fit/learn an acquisition function with an ansatz. Such an ansatz could take the form
            \begin{equation}
                y(v, w) = \sum_{r = 1}^k \sum_{i_1, ..., i_r = 0, ..., \Delta} w_{i_1, ..., i_r} v_{i_1} ... v_{i_r},
            \end{equation}
            where $k$ is the highest order of the ansatz and the variables $v_i$ depend on the continuous to discrete encoding. In terms of the indicator functions they are expressed as 
            \begin{equation}
                v_i = \sum_{\alpha} \alpha \delta^{\alpha}_{v_i}.
            \end{equation}
            Alternatively, one could write the ansatz as a function of the indicator functions alone instead of the variables
            \begin{equation}\label{eq:bbopt-2}
                y'(v, w) = \sum_{r = 1}^k \sum_{i_1, ..., i_r = 0, ..., \Delta} w'_{i_1, ..., i_r} \delta^{\alpha_{i_1}}_{v_{i_1}} ... \delta^{\alpha_{i_r}}_{v_{i_r}}.
            \end{equation}
            While the number of terms is roughly the same as before, this has the advantage that the energy scales in the cost function can be much lower. On the downside, one has to consider that usually for an optimization to be better than random sampling one needs the assumption that the function $f$ is well-behaved in some way (e.g. analytic). The formulation of the ansatz in Eq.~\ref{eq:bbopt-2} might not take advantage of this assumption in the same way as the first.
            
            Let ${w^* \equiv\{ w_{i_1, ..., i_r}^*\}}$ denote the fitted parameters of the ansatz. Then the cost function is simply $y(v, w^*)$ or $y'( v, w^*)$.
            \paragraph{Constraints}
            In this problem the only constraints that can appear are core terms. For the standard discretization no such term is necessary and for the random subspace coding we add Eq.~\eqref{def: rsc_core}.

            \paragraph{References}
            This problem can be found in Ref.~\cite{izawa2021continuous}.

            \subsubsection{Nurse scheduling with real parameters}

            In the following we will reexamine the nurse scheduling problem (see Sec.~\ref{sec:nurse_scheduling}) where the minimum workload $n_{\mathrm{min}, t}$ and the nurse power $p_i$ of a nurse $i$ are real-valued parameters. 
            The general structure of the cost function and the constraint will be the same as in Sec.~\ref{sec:nurse_scheduling} and when we choose to combine the cost function with the constraint of having a minimal workload per shift as in Eq.~\eqref{eq:ns-constraint} we do not have to change anything. However, there are several options when the constraint is implemented using the auxiliary variable $y_t$ defined in Eq.~\eqref{eq:ns-aux-var} and the conclusion reached by examining them might be useful for other problems. One possibility would be to use Eq.~\eqref{eq:ns-aux-var} with the real parameters
            $p_i$ and then $y_t$ will take only finitely many values 
            which we can label in some way. However, the problem is that the number of these values grows exponentially in the number of variables $N$. Alternatively, one might discretize the values of $y_t$ by splitting $y_{\mathrm{max}} \equiv \sum_{i=1}^N p_i$ into discrete segments. Again, this is not optimal as now the constraint that binds the auxiliary variable to its value (e.g., the penalty term in Eq.~\eqref{eq:ns-bind}) very likely is not exactly fulfilled and the remaining penalty introduces a bias in favor of configurations of $v_{i,t}$ where this penalty is lower but the cost function for the optimization might not be minimized.

            Thus, the natural choice is to simply discretize the parameters $n_{\mathrm{min}, t}, p_i$ and rescale them such that they take integer values. That is, we replace $p_i \rightarrow \tilde p_i \in \{1, ..., K \}$ where we set $K$ according to $K r = \underset{i}{\mathrm{max}}  \; p_i$ with the precision $r$ and we have $\tilde p_i r \leq p_i < (\tilde p_i +1)r$.

 		\subsection{Other problems } \label{ss:other_problems}
 		In this final category we include problems that do not fit into the previous classifications but constitute important use cases for quantum optimization. 
 		    
        \subsubsection{Syndrome decoding problem}
                
        \paragraph{Description}
        For an $[n, k]$ classical linear code (a code where $k$ logical bits are encoded in $n$ physical bits) the parity check matrix $H$ indicates whether a state $y$ of physical bits is a code word or has a non-vanishing error syndrome ${\eta= y H^T}$. Given such a syndrome, we want to decode it, i.e., find the most likely error with that syndrome, which is equivalent to solving
        \begin{equation}
            \underset{e \in \{0, 1\}^n, eH^T=\eta}{\text{argmin}} wt(e),
            \end{equation}
        where $wt$ denotes the Hamming weight and all arithmetic is mod 2.
                
        \paragraph{Cost function}
        There are two distinct ways to formulate the problem: Check-based and generator-based. In the generator-based approach we note that the generator matrix $G$ of the code satisfies $GH^T = 0$ and thus any logical word $u$ yields a solution to  $eH^T = \eta$ via $e = uG + v$ where $v$ is any state such that $vH^T = 0$ which can be found efficiently. Minimizing the weight of $uG + v$ leads to the cost function
        \begin{equation}
            f_G = \sum_{j = 1}^n (1-2v_j)  \left(1-2 \sum_{l=1}^k \delta^1_{u_l} G_{jl}\right).
        \end{equation}
        Note that the summation over $l$ is mod 2 but the rest of the equation is over $\mathbb{N}$.
        We rewrite this as 
        \begin{equation}\label{eq:check-reform}
            f_G = \sum_{j = 1}^n (1-2v_j) \prod_{l=1}^k s_l^{G_{jl}}
        \end{equation}
        according to Sec.~\ref{bb:mod2}.
                
        In the check-based approach we directly minimize deviations from $eH^T = \eta$ with the cost function
        \begin{equation}
            f_1 = \sum_{j = 1}^n (1-2\eta_j) \prod_{l=1}^k s_l^{H_{jl}}
        \end{equation}
        but we additionally have to penalize higher weight errors with the term
        \begin{equation}
            f_2 = \sum_{j = 1}^n \delta^1_{e_j}
        \end{equation}
        so we have
        \begin{equation}
            f_H \equiv c_1 f_1 + c_2 f_2
        \end{equation}
        with positive parameters $c_1/c_2$.
        \paragraph{Variables}
        The variables in the check-based formulation are the $n$ bits $e_i$ of the error $e$. In the generator-based formulation, $k$ bits $u_i$ of the logical word $u$ are defined. If we use the reformulation Eq.~\eqref{eq:check-reform}, these are replaced by the $k$ spin variables $s_l$. Note that the state $v$ is assumed to be given by an efficient classical calculation.
                
        \paragraph{Constraints}
        There are no hard constraints, any logical word $u$ and any physical state $e$ are valid.
                
        \paragraph{Resources}
        There are a number of interesting tradeoffs that can be found by analyzing the resources needed by both approaches.
        The check-based approach features a cost function with up to $(n-k)+n$ terms (for a non-degenerate check matrix with $n-k$ rows, there are $(n-k)$ terms from $f_1$ and $n$ terms from $f_2$) whereas in $f_G$ there are at most $n$ terms (the number of rows of $G$). The highest order of the variables that appear in these terms is for the generator-based formulation bounded by the number of rows of $G$ which is $k$. In general, this order can be up to $n$ (number of columns of $H$) for $f_H$ but for an important class of linear codes (low density parity check codes) the order would be bounded by the constant weight of the parity checks. This weight can be quite low but there is again a tradeoff because higher weights of the checks result in better encoding rates (i.e. for good low density parity check codes codes they increase the constant encoding rate $n/k \sim \alpha$).
        \paragraph{References}
        This problem was originally presented in Ref.~\cite{lai2022syndrome}.
                
        \subsubsection{$k$-SAT}\label{ss:k-sat}
		
        \paragraph{Description}
        A $k$-SAT problem instance consists of a boolean formula
        \begin{equation}
            f(x_1, ..., x_n) = \sigma_1(x_{1_1}, ..., x_{1_k}) ... \land \sigma_m(x_{m_1}, ..., x_{m_k})    
        \end{equation}
        in the conjunctive normal form (CNF), that is, $\sigma_i$ are disjunction clauses over $k$ literals $l$:
        \begin{equation}
            \sigma_i (x_{i_1}, ..., x_{i_k})= l_{i,1} \lor \dots \lor l_{i,k}, 
        \end{equation}
        where a literal $l_{i,k}$ is a variable $x_{i,k}$ or its negation $\neg x_{i,k}$.
        We want to find out if there exists an assignment of the variables that satisfies the formula.
            
        \paragraph{Variables}
        There are two strategies to express a $k$-SAT problem in a Hamiltonian formulation. First, one can use a Hamiltonian cost function based on violated clauses. In this case the variables are the assignments $x \in \{0, 1\}^n$. For the second method a graph is constructed from a $k$-SAT instance in CNF as follows. Each clause $\sigma_j$ will be a fully connected graph of $k$ variables $x_{i_1}, ..., x_{i_k}$. Connect two vertices from different clauses if they are negations of each other, i.e. $y_{i_l} = \neg y_{j_p}$ and solve the maximum independent set (MIS) problem for this graph~\cite{choi2010adiabatic}. If and only if this set has cardinality $m$, the SAT instance is satisfiable. The variables in this formulation are the $m\times k$ boolean variables $x_i$ which are 1 if the vertex is part of the independent set and 0 otherwise.
            
        \paragraph{Cost function}
        A clause-violation based cost function for a problem in CNF can be written as
        \begin{equation}
            f_C = \sum_j (1- \sigma_j(x))
        \end{equation}
         with (see Sec.~\ref{bb:bool})
         \begin{equation}
            \begin{aligned}
            \sigma_j &= l_{j_1} \lor l_{j_2} ... \lor l_{j_k} 
            \\ &= 1 -  \prod_{n=1}^k (1 - l_{j_n}).
            \end{aligned}
        \end{equation}
        A cost function for the MIS problem can be constructed from a term encouraging a higher cardinality of the independent set:
        \begin{equation}
            f_B = -b \sum_{j=1}^{m\times k} x_j.  
        \end{equation}

        \paragraph{Constraints}
        In the MIS formulation one has to enforce that there are no connections between members of the maximally independent set:
        \begin{equation}
            c = \sum_{(i,j) \in E} x_i x_j = 0.
        \end{equation} 
        If this constraint is implemented as an energy penalty $f_A = a c$, it should have a higher priority. The minimal cost of a spin flip (cf. Sec.~\ref{bb:prio}) from $f_A$ is $a(m-1)$ and in $f_B$ a spin flip could result in a maximal gain of $b$. Thus $a/b \gtrsim 1/(m-1)$.
            
        \paragraph{Resources}
        The cost function $f_C$ consists of up to $\mathcal{O}(2^k mk)$ terms with a maximal order of $k$ in the spin variables. In the alternative approach the order is only quadratic so it would naturally be in a QUBO formulation. However, the number of terms in $f_{\text{MIS}}$ can be as high as $\mathcal{O}(m^2 k^2)$ and thus scales worse in the number of clauses.

        \paragraph{References}
        This problem can be found in Ref.~\cite{choi2010adiabatic}.

    \subsubsection{Improvement of Matrix Multiplication}
    \paragraph{Description}
    Matrix multiplication is among the most used mathematical operations employed in informatics. For a streamlined presentation of the procedure we present the case of matrix multiplication over the real numbers. 
    
    Given a real $n\times m$ matrix $A$ and a real $m\times p$ matrix $B$ the usual algorithm for matrix multiplication requires $npm$ multiplication operations. However, there are algorithms using fewer resources, e.g. in Ref.~\cite{strassen1969gaussian}. Strassen showed that there is an algorithm for multiplying two $2\times 2$ matrices involving only $7$ multiplications at the expense of more addition operations which are computationally less costly. 
    
    In a recent paper~\cite{matrix} machine learning methods were used to identify new effective algorithms for matrix multiplication. They formulated the search for more effective algorithms (using less multiplication operations) as an optimization problem. We will show that there is an equivalent Ising-Hamlitonian optimization problem. 
    
    Multiplying matrices $A$ and $B$ can be stated in terms of an $nm\times mp\times np$-tensor $\mathcal{M}$. The usual algorithm for matrix multiplication corresponds to a sum decomposition of $\mathcal{M}$ into $nmp$-terms of the following form:
    \begin{equation}\label{eq: multiplication tensor}
        \mathcal{M}=\sum_{\alpha=0}^{nmp-1}\mathbf{u}^{(\alpha)}\otimes\mathbf{v}^{(\alpha)}\otimes \mathbf{w}^{(\alpha)}\, ,
    \end{equation}
    where $\mathbf{u}^{(\alpha)}$ is an $nm$-vector with binary entries for any $\alpha=0,\cdots, nmp$ and $\mathbf{v}^{(\alpha)}$, $\mathbf{w}^{(\alpha)}$ are binary $mp$- and $np$ vectors. How to multiply two matrices using the tensor Eq.~\eqref{eq: multiplication tensor} is described in Ref.~\cite[Algorithm~1]{matrix}. One maps matrices $A,B$ and $C$ to vectors of the respective length, e.g., an $n\times m$-matrix $A$ will be an $nm$-vector whose $k$-th entry is the component $A_{rs}$ for $k=rm+s$. Then one starts by computing an intermediate $nmp$-vector ${\mathbf{m}=(m_0,\cdots, m_{nmp-1})}$ which is computed by
    \begin{equation}
        m_i=\left(\sum_{j=0}^{nm-1}u^{(i)}_jA_j\right)\left(\sum_{k=0}^{mp-1}v_k^{(i)}B_i\right)
    \end{equation}
    for all ${i=0,\dots , nmp-1}$. In a second step, the components $\left\lbrace C_i\right\rbrace_{i=0,\dots, np-1}$ of the final matrix $C=AB$ are determined via
    \begin{equation}
        C_k=\sum_{r=0}^{nmp-1}w_k^{(r)}m_k\; ,
    \end{equation}
    for all $k{=0,\dots, np-1}$. Making matrix multiplication more efficient corresponds to finding sum decomposition of $\mathcal{M}$ with fewer terms (lower rank), i.e., we want to find vectors $\mathbf{a}^{(\beta)}$, $\mathbf{b}^{(\beta)}$, $\mathbf{c}^{(\beta)}$, where ${\beta=0,\dots , B-1}$ and ${B<nmp}$, such that 
    \begin{equation}
       \mathcal{M}=\sum_{\beta=0}^{B-1}\mathbf{a}^{(\beta)}\otimes\mathbf{b}^{(\beta)}\otimes \mathbf{c}^{(\beta)}\, 
    \end{equation}
    holds. Performing the above algorithm with this  shorter tensor decomposition corresponds to a matrix multiplication algorithm involving fewer scalar multiplications. Thus the new algorithm is more efficient. In general, we can allow the vectors $$\left\lbrace \mathbf{a}^{(\beta)}, \mathbf{b}^{(\beta)}, \mathbf{c}^{(\beta)}\right\rbrace_{\beta=1,\cdots , B-1}$$ to have real entries. To ease our lives we restrict ourselves to entries in the finite set $K\coloneqq \left\lbrace -k,-k+1,\dots, k-1,k \right\rbrace$ with $k\in \mathbb{N}$ and Ref.~\cite{matrix} reports promising results for $k=2$. Note that the tensor $\mathcal{M}$ has binary entries though. Allowing more values for the decomposition vectors just increases the search space for possible solutions. 
   \paragraph{Variables} The variables of the problem are given by the entries of the vectors in a sum decomposition of $\mathcal{M}$. That is, there are vectors of variables $$\left\lbrace \mathbf{a}^{(\alpha)}\right\rbrace_{\alpha=1,\dots, nmp-1}$$, where every entry $a_i^{(\alpha)}$ of ${\mathbf{a}^{(\alpha)}=(a_0^{(\alpha)},\dots, a_{nm-1}^{(\alpha)})}$ is a discrete variable with range $K$. Similarly, we have vectors of variables ${\left\lbrace \mathbf{b}^{(\alpha)}_{\alpha=0,\dots,nmp-1}\right\rbrace}$, with ${\mathbf{b}^{(\alpha)}=(b_0^{(\alpha)},\dots, b_{mp}^{(\alpha)})}$, where $b_i^{(\alpha)}$ has range $K$ and vectors of variables ${\left\lbrace \mathbf{c}^{(\alpha)}\right\rbrace_{\alpha=0,\dots, nmp-1}}$ with ${\mathbf{c}^{(\alpha)}=(c_0^{(\alpha)},\dots, c_{np-1}^{(\alpha)})}$ and $c_i^{(\alpha)}$ has again range $K$.
   
   \paragraph{Cost function}
   The cost function of the problem is given by 
   \begin{equation}
       f_\mathcal{M}\coloneqq \sum_{\alpha=0}^{nmp-1}\left(1-\prod_{i,j,k}\delta_{a_i^{(\alpha)}}^0\delta_{b_j^{(\alpha)}}^0\delta_{c_k^{(\alpha)}}^0\right).
   \end{equation}
   Note that for fixed $\alpha$, the term
   \begin{equation}
       \left(1-\prod_{i,j,k}\delta_{a_i^{(\alpha)}}^0\delta_{b_j^{(\alpha)}}^0\delta_{c_k^{(\alpha)}}^0\right)
   \end{equation}
   vanishes if and only if all three vectors $\mathbf{a}^{(\alpha)}$, $\mathbf{b}^{(\alpha)}$ and $\mathbf{c}^{(\alpha)}$ vanish. This cost function thus penalizes higher rank in the sum decomposition~\eqref{eq: multiplication tensor} and therefore the groundstate will represent a more effective multiplication algorithm. 
   \paragraph{Constraints} We have to impose constraints such that the resulting tensor decomposition really coincides with the matrix multiplication tensor $\mathcal{M}$. Thus there are constraints
   \begin{equation}
       \mathcal{M}_{ijk}=\sum_{\alpha=0}^{nmp-1}\mathbf{a}_i^{(\alpha)}\mathbf{b}^{(\alpha)}_{j}\mathbf{c}^{(\alpha)}_{k}
   \end{equation}
   for any ${i=0,\dots, nm-1}$, ${j=0,\dots, mp-1}$ and ${k=0,\dots,np-1}$. 
   
   Note that the number of constraints as well as the number of terms in the cost function grows fast in the dimension of the matrices to be multiplied.
   
    \section{Summary of building blocks}\label{sec:bb}
 	Here we summarize parts of cost functions and techniques that are used as reoccurring building blocks for the problems in this library.

        \subsection{Auxiliary functions}
        The simplest class of building blocks are auxiliary scalar functions of multiple variables that can be used directly in cost functions or constraints via penalties.
  
		\subsubsection{Element counting}\label{bb:ele_cnt}
		One of the most common building blocks is the function $t_a$ that simply counts the number of variables $v_i$ with a given value $a$. In terms of the value indicator functions we have
		\begin{equation}
		    t_a = \sum_{i} \delta^a_{v_i}.
		\end{equation}
		It might be useful to introduce $t_a$ as additional variables. In that case one has to bind it to its desired value with the constraints
		\begin{equation}
		    c = \left(t_a - \sum_i \delta^a_{v_i} \right)^2.
		\end{equation}

        \subsubsection{Step function}\label{bb:step_func}
        Step functions $\Theta(v-w)$ can be constructed as
        \begin{equation}\label{eq:theta}
            \Theta(v-w) = \sum_{\beta \geq \alpha}^K \delta^{\beta}_v \delta^{\alpha}_{w}=\begin{cases} 
            1 \, &\text{if}\, v\geq w\\
            0\, &\text{if}\, v< w
            \end{cases},
        \end{equation}
        with $K$ being the maximum value that the $v,w$ can take on. Step functions can be used to penalize configurations where $v\geq w$. 
        
        \subsubsection{Minimizing the maximum element of a set}\label{bb:min_maximum}
        Step functions are particularly useful for minimizing the maximum value of a set $\{v_i\}$. Given an auxiliary variable $l$, we guarantee that $l\geq v_i,$ for all $i$ with the penalization
		\begin{equation}\label{eq:min-max}
		    f = 1- \prod_{i=1}^n \Theta(l-v_i)=\begin{cases} 
            1 \, &\text{if}\; l \geq v_i\; \forall i\\
            0\, &\text{if}\; \exists\; i:\;v_i>l,
            \end{cases}
		\end{equation}
        which increases the energy if $l$ is smaller than any $v_i$. The maximum value of $\{v_i\}$ can be minimized by adding to the cost function of the problem the value of $l$. In that case we also have to multiply the term from Eq.~\eqref{eq:min-max} by the maximum value that $l$ can take in order to avoid trading off the penalty.

        \subsubsection{Compare variables} \label{bb:compare variables}
        Given two variables $v,w\in[1,K]$, the following term indicates if $v$ and $w$ are equal:
        \begin{equation} \label{eq:compare_variables_indicator}
				        \delta(v - w) = \sum_{\alpha=1}^{K} \delta_{v}^{\alpha} \delta_{w}^{\alpha}=\begin{cases}1\; \mathrm{if}\, v=w
			        \\ 0\; \mathrm{if}\, v\neq w.
            \end{cases}
	    \end{equation}
        If we want to check if $v>w$, then we use the step function Eq.~\eqref{eq:theta}.

        \subsection{Constraints}
        Here we present the special case of functions of variables where the groundstate fulfills useful constraints. These naturally serve as building blocks for enforcing constraints via penalties.
        
		\subsubsection{All (connected) variables are different}\label{bb:all_different}
			If we have a set of variables $\lbrace v_i\in [1,K]\rbrace_{i=1}^N$ and two variables $v_i, v_j$ are connected when the entry $A_{ij}$ of the adjacency matrix $A$ is 1, the following term has a minimum when $v_i\neq v_j$ for all connected $i\neq j$:
			\begin{equation}
			    \begin{aligned}
			        c &= \sum_{i\neq j} \delta(v_i-v_j) A_{ij}
			        \\&= \sum_{i\neq j} \sum_{\alpha=1}^{K} \delta_{v_i}^{\alpha}\delta_{v_j}^{\alpha} A_{ij}
			    \end{aligned}
			\end{equation}
			The minimum value of $c$ is zero, and it is only possible if and only if there is no connected pair $i,j$ such that ${v_i=v_j}$. For all-to-all connectivity one can use this building block to enforce that all variables are different. If ${K=N}$, the condition ${v_i\neq v_j}$ for all $i\neq j$ is then equivalent to asking that each of the $K$ possible values is reached by a variable.
		
		\subsubsection{Value $\alpha$ is used} \label{bb:alpha_is_used}
		    Given a set of $N$ variables $v_i$, we want to know if at least one variable is taking the value $\alpha$. This is done by the term
		    \begin{equation}
                u_{\alpha} =\prod_{i=1}^{N} \left(1-\delta_{v_i}^{\alpha} \right)=
                \begin{cases}
                0\qquad\mathrm{if}\; \exists\, v_i=\alpha
                \\ 1 \qquad\mathrm{if}\; \nexists\, v_i=\alpha.
                \end{cases}
		    \end{equation}

		\subsubsection{Inequalities}\label{bb:ineq}
		\paragraph{Inequalities of a single variable}
		    If a discrete variable $v_i$ which can take values in $1, ..., K$ is subject to an inequality $v_i \leq a'$ one can enforce this with a energy penalization for all values that do not satisfy the inequality
		    \begin{equation}
		        c = \sum_{K>a>a'} \delta_{v_i}^a.
		    \end{equation}
		    It is also possible to have a weighted penalty, e.g.
		    \begin{equation}
		        c = \sum_{K\geq a>a'} a \delta_{v_i}^a,
		    \end{equation}
		    which might be useful if the inequality is not a hard constraint and more severe violations should be penalized more. That option comes with the drawback of introducing in general higher energy scales in the system, especially if $K$ is large, which might decrease the relative energy gap.

		\paragraph{Inequality constraints}		    
		    If the problem is restricted by an inequality constraint:
		    \begin{equation}
		        c(v_1,\dots,v_N)<K,
		    \end{equation}
		    it is convenient to define an auxiliary variable $y<K$ and impose the constraint:
		    \begin{equation} \label{eq:bb_ineq_bound}
		        \left( y - c(v_1,\dots,v_N)\right)^2 = 0,
		    \end{equation}
	        so $c$ is bound to be equal to some value of $y$, and the only possible values for $y$ are those that satisfy the inequality. 
	        
	        If the auxiliary variable $y$ is expressed in binary or gray encoding, then Eq.~\eqref{eq:bb_ineq_bound} must be modified when $2^n <K< 2^{n+1}$ for some $n\in \mathbb{N}$, 
		    \begin{equation} 
		        \left( y- c(v_1,\dots,v_N) -2^{n+1} + K\right)^2 = 0,
		    \end{equation}
		    which ensures $c(v_1,\dots,v_N)<K$ if $y = 0,\dots, 2^{n+1}-1$.

        \subsubsection{Constraint preserving driver}
        When annealing inspired algorithms like quantum annealing or QAOA are used, an alternative to add penalty terms to the cost function that penalize states which do not satisfy constraint is to start the algorithm in a state that fulfills the constraints and use driver Hamiltonians that do not lead out of the constraint-fulfilling subspace ~\cite{hadfield2019quantum, Ender2022modular}. As explained in Sec.~\ref{ss: constraints}, we demand that the driver Hamiltonian commutes with the operator generating the constraints and that it reaches the entire valid search space. 
        Arbitrary polynomial constraints $c$ can for example be handled by using the Parity mapping. It brings constraints to the form $\sum_{\mathbf u \in c} g_{\mathbf u} \sigma^{(\mathbf u)}_z$. Starting in a constraint-fulfilling state and employing constraint depended flip-flop terms constructed from $\sigma_+, \sigma_-$ operators
        as driver terms on the mapped spins~\cite{drieb2021encoding} automatically enforces the constraints.

        \subsection{Problem specific representations}
        For selected problems that are widely applicable, we demonstrate useful techniques for mapping them to cost functions.
        
		\subsubsection{Modulo 2 linear programming}\label{bb:mod2}
            For the set of linear equations 
            \begin{equation}\label{eq:lin-prog}
                x A = y,
            \end{equation}
            where $A \in \mathbb{F}_2^{l \times n}, y \in \mathbb{F}_2^n$ are given, we want to solve for $x \in \mathbb{F}_2^l$. The cost function 
            \begin{equation}
                f = \sum_{i = 1}^n (1-2y_i)  \left(1 - \sum_{j=1}^l 2 x_j A_{ji}\right)
            \end{equation}
            minimizes the Hamming distance between $xA$ and $y$ and thus the ground state represents a solution. If we consider the second factor for fixed $i$, we notice that it counts the number of $1$s in $x$ (mod 2) where $A_{ji}$ does not vanish at the corresponding index. When acting with
            \begin{equation}
                H_i =  \prod_{j=1}^l \sigma_{z,j}^{A_{ji}}
            \end{equation}
            on $|x \rangle$ we find the same result and thus the cost function 
            \begin{equation}
                f = \sum_{i = 1}^n (1-2y_i) \prod_{j=1}^l s_j^{A_{ji}},
            \end{equation}
           with spin variables $s_j$ has the solution to Eq.~\eqref{eq:lin-prog} as its ground state.
            
        \subsubsection{Representation of boolean functions}\label{bb:bool}
        Given a boolean function $f: \{0,1\}^n \rightarrow \{0,1\}$ we want to express it as a cost function in terms of the $n$ boolean variables. This is hard in general~\cite{hadfield2021representation}, but for (combinations of) local boolean functions there are simple expressions. In particular we have, e.g., \begin{align}
            \neg x_1 = & 1-x_1 \\
            x_1 \land x_2 \land ... \land x_n = & \prod_{i=1}^n x_i \\
            x_1 \lor x_2 \lor ... \lor x_n = & 1-\prod_{i=1}^n (1-x_i) \\
            (x_1 \rightarrow x_2) = & 1-x_1 + x_1 x_2 \\
            \mathrm{XOR}(x_1, x_2) \equiv x_1 + x_2 \;\mathrm{mod } 2 = & x_1 + x_2 -2x_1 x_2.  
            \end{align}

        It is always possible to convert a $k$-SAT instance to 3-SAT and more generally any boolean formula to a 3-SAT in conjunctive normal form with the Tseytin transformation~\cite{tseitin1983complexity} with only a linear overhead in the size of the formula.
        
        Note that for the purpose of encoding optimization problems one might use different expressions that only need to coincide (up to a constant shift) with those shown here for the ground state/solution to the problem at hand. For example, if we are interested in a satisfying assignment for $x_1 \land ... \land x_n$ we might formulate the cost function in two different ways that both have $x_1 = ... =x_n =1$ as their unique ground state; namely $f = -\prod_{i=1}^n x_i$  and $f = \sum_{i=1}^n (1-x_i)$. The two corresponding spin Hamiltonians will have vastly different properties, as the first one will have, when expressed in spin variables, $2^n$ terms with up to $n$-th order interactions, whereas the second one has $n$ linear terms. Additionally, configurations which are closer to a fulfilling assignment, i.e. have more of the $x_i$ equal to one, have a lower cost in the second option which is not the case for the first option. 
        Note that for $x_1 \lor ... \lor x_n$ there is no similar trick as we want to penalize exactly one configuration.

        \subsection{Meta optimization}
        Finally, we present several options for choices that arise in the construction of cost functions. This includes meta parameters like coefficients of building blocks, but also reoccurring methods and techniques in dealing with optimization problems.  

        \subsubsection{Auxiliary variables}\label{bb:helper_fct}
	    It is often useful to combine information about a set of variables $v_i$ into one auxiliary variable $y$ that is then used in other parts of the cost function. Examples include the element counting building block or the maximal value of a set of variables where we showed a way to bind the value of an auxiliary variable to this function of the set of variables. If the function $f$ can be expressed algebraically in terms of the variable values/indicator functions (as a finite polynomial) the natural way to do this is via adding the constraint term
	    \begin{equation}
	       c \left(y - f(v_i) \right)^2 
	    \end{equation}
		with an appropriate coefficient $c$. To avoid the downsides of introducing constraints, there is an alternative route that might be preferred in certain cases. This alternative consists in simply using the variable indicator $\delta^{\alpha}_{y(v_i, \delta^{\beta}_{v_i})}$ and expressing the variable indicator function in terms of spin/binary variables according to a chosen encoding (one does not even have to use the same encoding for $y$ and the $v_i$). As an example, let us consider the auxiliary variable $\tau_{m,t}$ in the machine scheduling problem \ref{sec:machine-scheduling} for which we need to express $\delta_{\tau_{m,t}}^{\alpha}$. For simplicity, we choose the one-hot encoding for $v_{m,t}$ and $\tau_{m,t}$, leading to 
		\begin{equation}
		    \delta^{\alpha}_{\tau_{m,t}} = x_{\tau_{m,t}, \alpha} = \begin{cases}(1-x_{v_{m,t}, 0}) & \mathrm{if}\, \alpha = t \neq 0
			        \\ x_{v_{m,t}, 0} & \mathrm{if}\, t = 0 \\
			        0  &\mathrm{else}, 
			        \end{cases}
		\end{equation}
		where the $x_{v_{m,t}, \beta}$ are the binary variables of the one-hot encoding of $v_{m,t}$.
        
        \subsubsection{Prioritization of cost function terms}\label{bb:prio}
            If a cost function is constructed out of multiple terms, 
            \begin{equation}
            f = a_1 f_1 + a_2 f_2,    
            \end{equation}
             one often wants to prioritize one term over another, e.g., if the first term encodes a hard constraint.   
             One strategy to ensure that $f_1$ is not ``traded off" against $f_2$ is to evaluate the minimal cost $\Delta_1$ in $f_1$ and the maximal gain $\Delta_2$ in $f_2$ from flipping one spin. 
             It can be more efficient to do this independently for both terms and assume a state close to an optimum. Then the coefficients can be set according to $c_1 \Delta_1 \gtrsim c_2 \Delta_2$. 
             In general, this will depend on the encoding for two reasons; first, for some encodings one has to add core terms to the cost function which have to be taken into account for the prioritization (they usually have the highest priority). Furthermore, in some encodings a single spin flip can cause the value of a variable to change by more than one.\footnote{E.g. in the binary encoding a spin flip can cause a variable shift of up to $K/2$ if the variables take values $1, ..., K$. This is the main motivation to use modifications like the Gray encoding.}
             Nonetheless, it can be an efficient heuristic to compare the ``costs" and ``gains" introduced above for pairs of cost function terms already in the encoding independent formulation by evaluating them for single variable changes by one and thereby fix their relative coefficients. Then one only has to fix the coefficient for the core term after the encoding is chosen.
             
        \subsubsection{Problem conversion}
        Many decision problems associated to discrete optimization problems are NP-complete, meaning that one can always map them to any other NP-complete problem with only a polynomial overhead of classical runtime in the system size. Therefore, a new problem for which one does not have a cost function formulation could be classically mapped to another problem where such a formulation is at hand. However, since quantum algorithms are expected to deliver only a polynomial advantage for general NP-hard problems, it is advisable to carefully analyze the overhead. It is also possible that there are trade-offs as in the example $k$-SAT in Sec.~\ref{ss:k-sat}, where in the original formulation the order of terms in the cost function is $k$ while in the MIS formulation we naturally have a QUBO problem where the number of terms scales worse in the number of clauses.

    \section{Conclusion and Outlook}\label{sec:conclusion}

 In this paper we have collected and elaborated on a wide variety of optimization problems, which 
 can be formulated as a spin Hamiltonian, allowing them to be solved on a quantum computer.
 We first presented these optimization problems in an encoding independent fashion. Only after choosing an encoding for the problem's variables, do we obtain a spin Hamiltonian ready to be fed into algorithms such as quantum annealing or QAOA. Representing an optimization problem in terms of a spin Hamiltonian already comprises some choices, one of which is an encoding. Ultimately, this freedom arises from the fact that the solution to the problem is only encoded in the ground state of the Hamiltonian and the excited states are not fixed. 

 This is best illustrated with an example from the representation of the one-hot core term in 
 Sec.~\ref{ss:onehot} where two alternatives were given to enforce the one-hot condition.
 If we were restricted to convert problems to a QUBO form, the choice with lower order of interaction would be superior but with the Parity mapping and architecture, higher order interaction terms can be handled better and thus the Hamiltonian with fewer terms might be preferred.

We have thus highlighted that obtaining the spin Hamiltonian for an optimization problem involves several choices. Rather than a burden, we consider this a feature. It leaves room to improve the performance of quantum algorithms by making different choices. This is the second main contribution of this paper; we identified common, reoccurring building blocks in the formulation of  optimization problems. 
 These building blocks, like encodings, parts of cost functions or different choices for imposing constraints, constitute a construction kit for optimization problems. This paves the way to an automated preprocessing algorithm picking the best choices among the construction kit to facilitate the search for optimally performing quantum algorithms. 
 
 Finding the optimal spin Hamiltonian for a given hardware platform for a quantum computer is in itself an important problem. Optimal spin Hamiltonians for different hardware platforms are very likely to differ, thus it is important to have a procedure capable of tailoring a problem to any given platform. Our hardware agnostic approach is a first key step towards this direction.

 In Sec.~\ref{sec:using-lib} we also demonstrated for two examples of how to use the building blocks and make choices when constructing the spin Hamiltonian. These examples can be used as a template to solve customized optimization problems in a systematic way. 

The next step in this program towards automation is some explicit testing for selected problems. In particular, we plan to benchmark different encodings for variables. 
Ultimately, the testing and benchmark results will be incorporated in a pipeline to automate the search for Hamiltonian formulations leading to better quantum algorithm performance.

\section{Acknowledgements}
The authors thank Dr. Kaonan Micadei for fruitful discussions. Work was supported by the
  Austrian Science Fund (FWF) through a START grant under Project No. Y1067-N27, the SFB BeyondC Project No. F7108-N38,  QuantERA II Programme under Grant Agreement No. 101017733, the Federal Ministry for Economic Affairs and Climate Action through project QuaST, and the Federal Ministry of Education and Research on the basis of a decision by the German Bundestag. 
\bibliography{bibliography}

\end{document}